\documentclass[11pt,a4paper]{article}
\usepackage{jinstpub}
\newcommand{\caloc}[0]{\textsc{CaloClouds}}
\newcommand{\ccthree}[0]{\textsc{CaloClouds3}}
\newcommand{\cctwo}[0]{\textsc{CaloClouds2}}
\newcommand{\gfour}[0]{\textsc{Geant4}}
\newcommand{\showerf}[0]{\textsc{ShowerFlow}}
\newcommand{\ddsim}[0]{\textsc{DDsim}}
\newcommand{\ddml}[0]{\textsc{DDML}}
\newcommand{\ddforhep}[0]{\textsc{DD4hep}}
\newcommand{\keyforhep}[0]{\textsc{Key4Hep}}
\newcommand{\torchscript}[0]{\textsc{TorchScript}}

\title{\ccthree{}: Ultra-Fast Geometry-Independent Highly-Granular Calorimeter Simulation}

\author[a,b]{Thorsten Buss}
\author[a,1]{Henry Day-Hall%
\note{Corresponding author. Authorship is alphabetical.}}
\author[a]{Frank Gaede}
\author[b]{Gregor Kasieczka}
\author[a]{Katja Krüger}
\author[a]{Anatolii Korol}
\author[a]{Thomas Madlener}
\author[c]{Peter McKeown}
\author[b]{Martina Mozzanica}
\author[b]{Lorenzo Valente}
\affiliation[a]{Deutsches Elektronen-Synchrotron DESY, Notkestr. 85, 22607 Hamburg, Germany}
\affiliation[b]{Institut für Experimentalphysik, Universität Hamburg, Luruper Chaussee 149, 22761 Hamburg, Germany}
\affiliation[c]{CERN, 1211 Geneva 23, Switzerland}
\emailAdd{henry.day-hall@desy.de}
\keywords{Detector modelling and simulations I, Simulation methods and programs, Calorimeter methods}

\arxivnumber{2511.01460}

\usepackage{siunitx}

\usepackage{amsmath}

\usepackage{gensymb}

\usepackage{graphicx}

\usepackage{hyperref}

\usepackage[toc,page]{appendix}

\usepackage{xcolor}

\usepackage[normalem]{ulem}

\usepackage{makecell}

\abstract{
    We present \ccthree{}, a model for the fast simulation of photon showers in the barrel of a high granularity detector.
    This iteration demonstrates for the first time how a pointcloud model can employ angular conditioning to replicate photons at all incident angles. 
    Showers produced by this model can be used across the whole detector barrel, due to specially produced position agnostic training data.
    With this flexibility, the model is usable in a full simulation and reconstruction chain, which offers a further handle for evaluating physics performance of the model.
    As inference time is a crucial consideration for a generative model, the 
    pre-processing and hyperparameters are aggressively optimised, achieving a speed up factor of two orders of magnitude over \gfour{} at inference.
}

\begin{document}
\maketitle

\section{Introduction and Motivation}

Comparison between events observed in particle colliders and predictions made by theoretical models is an inverse problem.
We wish to assess the likelihood of the theory given the observations, but it is only possible to simulate samples of likely observations, given the theory.
Ensuring the statistical uncertainly of the samples does not dominate the analysis requires a vast quantity of simulation.
In experimental particle physics the detailed simulation of the response to particles in calorimeter detectors is the single most resource demanding step of such large Monte Carlo simulations.
Generating this with the computational resources available requires the use of fast simulation methods, such as generative Machine Learning (ML) methods~\cite{WallClock}. Some of those include generative adversarial networks (GANs)~\cite{Paganini:2017hrr, Paganini:2017dwg, deOliveira:2017rwa, Erdmann:2018kuh,
Musella:2018rdi, Erdmann:2018jxd, Belayneh:2019vyx, Butter:2020qhk, Javurkova:2021kms, Bieringer:2022cbs, Hashemi:2023ruu, 2024_atlas}, 
variational autoencoders (VAEs) and their variants
\cite{Buhmann:2020pmy, Buhmann:2021lxj, Buhmann:2021caf, 2024_atlas,Cresswell:2022tof,Diefenbacher:2023prl, hashemi2024deepgenerativemodelsultrahigh},
generative pre-trained transformer (GPT) style models~\cite{Birk:2025wai}, 
normalizing flows and diffusion \mbox{models ~\cite{sohldickstein2015deep,
song2020generative_estimatingGradients,
song2020improved_technieques_for_sorebased_geneerative, ho2020denoising,
song2021scorebased_generativemodelling, Mikuni:2022xry_CaloScore,
Acosta:2023zik, Mikuni:2023tqg,
Amram:2023onf, Chen:2021gdz, Krause:2021ilc, 
Krause:2021wez, schnake2022_pointFlow, Krause:2022jna,
Diefenbacher:2023prl, 
Xu:2023xdc, Buckley:2023daw,Ernst:2023qvn, OmanaKuttan:2024mwr,Favaro:2024rle, Brehmer:2024yqw, Buss:2025cyw, Raikwar:2025fky}}.

\begin{figure}[ht]
    \centering
    \includegraphics[width=0.8\textwidth]{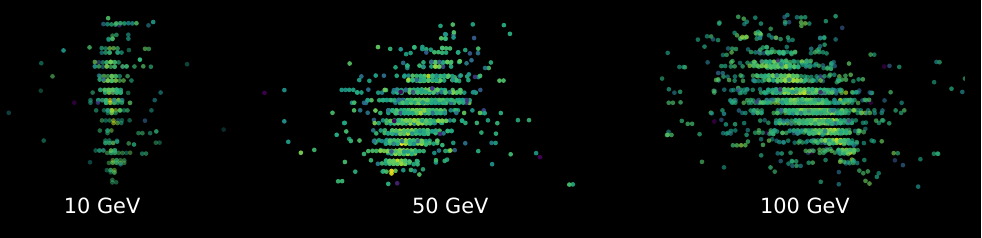}
    \caption{
        Photon showers, with varying incident angle, in the top barrel segment of the ILD SiECAL as simulated by \gfour{}.
        The incident energy of the photon gun is indicated beneath each shower.
    }
\label{fig:ExampleShower}
\end{figure}

In this paper, we present the latest of a series of models designed for the fast simulation of photon showers in a high granularity calorimeter, such as those depicted in figure~\ref{fig:ExampleShower}.
This is the latest of the \caloc{} series of models, which represent the showers to be modeled as pointclouds.
Two ML techniques are used; a normalising flow combined with a diffusion model\footnote{The code is avaliable at \url{https://github.com/FLC-QU-hep/CaloClouds-3}}.
For a review covering other possible architectures, see~\cite{taxonomic_review}.
For a comparison of recent models, see the companion paper, a first full physics benchmark for highly granular calorimeter surrogates~\cite{benchmarks} and results of the CaloChallange~\cite{CaloChallange}.

\gfour{}~\cite{Geant4} is a particle transport Monte Carlo toolkit for the detailed simulation of the interaction of particles with matter based on a detailed modelling of the underlying physics. It is used by the majority of experiments in particle physics but requires large compute resources, especially in calorimeter simulation.
Fast simulation aims to closely approximate the output of \gfour{}, at greater speed than is possible with the detailed Monte Carlo techniques employed by \gfour{}.

This paper describes the development of the current model from its predecessor.
We aim to both justify the conclusion that the model is an optimal example of its architecture, and provide perspectives useful to optimising other models.


The previous model, \cctwo{}, was only capable of simulating photons incident at a single angle and position.
Incident angle conditioning, along with location agnostic training data,
allows a single model to be used for all photons entering the calorimeter,
which simplifies training, and potentially occupies less memory at inference time.
The octagonal barrel shape and magnetic field, which is aligned with
the beam axis, break the rotational symmetry about the shower axis,
therefore the model must be conditioned on both polar (\(\theta\)) and azimuthal (\(\phi\)) directions separately.

Clearly this upgrade is needed for the model to be used in realistic simulation, but it also helps to judge the fidelity with which the physics is modelled.
Evaluating the success of a generative model is often challenging.
We know the model must replicate the shower development of \gfour{}, but it is difficult to decide on an acceptable precision.
Even \gfour{} produces some changes between versions~\cite{g4_verson_comp}, and perfectly replicating \gfour{} would likely be as slow as \gfour{}.
With angular conditioning, \ccthree{} is viable to be used in a realistic detector simulation;
full \gfour{} simulation is performed until a photon reaches the calorimeter. Assuming the photon satisfies the appropriate kinematic and geometrical constraints, it is then removed from the detailed full simulation, and replaced by the corresponding output of \ccthree{}.
We then run a standard reconstruction on these simulated events, allowing for realistic physics benchmarking of the ML model and thereby providing the ultimate test of the model
performance.

While inference speed for \cctwo{} was already \(\approx 40\) times faster than \gfour{} for \(100~\si{GeV}\) photons, the model hyperparameters were not chosen in a systematic manner, and were inspired by choices made in other domains.
Like many ML tools,  diffusion models were originally developed for use on natural images~\cite{originalDiffusion}.
Normalising flows were initially proposed in 2013~\cite{NormFlowStat}, but without the involvement of neural networks. Later variants that did use neural networks popularised the method~\cite{NormFlow1,NICE}, and were also developed for image processing. 
This has influenced standard hyperparameter choices, so it is likely that a fundamentally different task like ours might benefit from significant changes to these choices.

In section~\ref{sec:dataset} we describe the example detector we have chosen to demonstrate our architecture on, and the dataset generation is detailed. In section~\ref{sec:model} the new \ccthree{} model is defined. In sections~\ref{sec:accuracy} and~\ref{sec:reco} we compare physics accuracy between the previous iteration, \cctwo{}, and the new model, \ccthree{}. After this, a snapshot of the newly attained capacity for full reconstruction is offered in section~\ref{sec:reco}, however this is covered in more depth in a companion paper~\cite{benchmarks}.
In the penultimate section, section~\ref{sec:speed}, the improvements in inference speed between \cctwo{} and \ccthree{} are demonstrated.
Finally, we draw the conclusions in section~\ref{sec:conclusions}.

\section{Dataset}\label{sec:dataset}

While the techniques and model architecture described here is expected to be a good solution for photon simulation in any high granularity calorimeter, we require a concrete example for training and testing.
First, section~\ref{sec:detector} specifies the example detector chosen to generate the data.
Then section~\ref{sec:preprocessing} describes the preprocessing used to optimally interface the data and the model.

\subsection{Detector}\label{sec:detector}
The International Large Detector (ILD)~\cite{ILDrep} is chosen as an example, as it has a calorimeter typical of a Higgs Factory detector.
Higgs factories are designed to do precision physics, colliding fundamental particles (often electron-positron pairs) at energies in a tight range around the di-Higgs production mass.
The ILD is designed with a normal range of sub-detectors; a vertex detector, a tracking detector, a time projection chamber, calorimeters and a muon system.
However, in this study we are focused only on the calorimeters.
Specifically, we consider the silicon variant of the electromagnetic calorimeter (denoted SiECAL in the ILD report~\cite{ILDrep}, see section 5.1.2.4).
Overall, the barrel is constructed of 8 flat segments, forming an octagon.
As all segments have rotational symmetry, all simulated data is focused on the top segment.
Within each segment are 30 sensitive layers,
composed of active Si-cells of \(5\times5\)mm, with a thickness of \(0.5\)mm, interspersed at every other layer with passive Tungsten absorbers.
The Tungsten absorbers in the first 20 layers are \(2.1~\si{mm}\) thick, while the final 10 layers have thicker \(4.2~\si{mm}\) absorbers.
Such a high granularity calorimeter would provide sufficient detail to accurately reconstruct photons as low as \(0.1~\si{GeV}\), and while photons of energy \(\lesssim 10~\si{GeV}\) do not take so long to  generate with \gfour{}, replacing the higher energies with a faster technique would be essential.
A magnetic field of \(3.5~\si{T}\) parallel to the beam encompass the detector, and the curvature this causes in charged particles is seen in the shower development.

\subsection{Generation and preprocessing}\label{sec:preprocessing}
Photons between \(1\) and \(127~\si{GeV}\) were generated using \gfour{},
a slightly expanded range from the energies considered in \cctwo{}.
\(127~\si{GeV}\) has been chosen as it is just over half of the initial collision energy planned at the ILD~\cite{ILDrep}.
In the framework used here, a fast model is only utilised (triggered) once a photon in an appropriate energy range reaches the surface of the calorimeter, therefore any photons that start showering before the surface of the calorimeter will not be simulated with the fast simulation model\footnote{Potentially, some secondary photons from the shower of an early showering photon could reach the calorimeter surface with sufficient energy to trigger the fast simulation, but this would anyway have the same physics as a low energy primary photon, and so is covered by the dataset.}.
Consequently it is important to generate the datasets used to train the model without any pollution from early showering photons\footnote{For further description of how the model is triggered in the context of a complete event simulation, see section~\ref{sec:CC3arch}.}.
To avoid any early showering, the photons in the training set are inserted into the simulation at the surface of the calorimeter.
The angle of each photon is consistent with having originated from the Interaction Point (IP),
having \(30<\theta<150\) and \(80<\phi<110\), sampled to have an even density on the upper barrel segment. %
\gfour{} then records the hits of the shower in the sensitive layers.
The training data includes \(3\times 10^6\) events, with a test set of \(7\times10^4\) which was held back.

As the objective is to use the same shower model for the entire detector, the training data must not contain specific information about support structures or dead zones.
Additionally, the edges of each segment of the hexagonal barrel join at a \(135\degree \) angle, meaning each layer starts at differing positions and calorimeter tiles are somewhat staggered between layers. 
We would prefer that the shower model does not learn these localised effects. For this reason, the detector is regularised; all supporting structure that interrupts the plane of the layers is removed, and the cells are aligned to form an identical grid in each layer. 
The shower model is then trained on this, location agnostic, regularised geometry.
At inference time, the effects of the detector geometry at the relevant position are recovered by projecting the showers produced by the model back into a realistic detector map, as described in the next subsection~\ref{sec:CC3arch}. 

A local coordinate system for the shower is defined with the \(z\) axis going through the sensitive layers,
both the \(x\) and \(y\) coordinates are parallel to the surface of the layers, with the \(x\) coordinate also parallel to the beam direction.

The centre of each layer is shifted in the \(x\) and \(y\) directions to remove the shower's tilt.
This shift corresponds to the rotation that would align the incident photon with the \(z\) axis,
so the size of the shift is calculated from the depth of the layer \(d_\mathrm{layer}\) and the momentum of the incident particle \(\vec{p}_\mathrm{inc}\);
in the \(x\) direction, the shift is \(\Delta x = d_\mathrm{layer}\tan(\sin^{-1}(\vec{p}_{\mathrm{inc},x}/|\vec{p}_\mathrm{inc}|))\), and likewise for the \(y\) direction.
To first approximation, the shifted showers are now a similar oblong shape, somewhat akin to one created by an upward moving particle.
This preprocessing helps the diffusion model to learn the details of backscattering, rather than being forced to focus on getting the gross shape right, which is known from the direction of the incident particle.

Finally, all 30 layers of the detector are trimmed to a box of \(100\times100\) cells,  
cutting to \(\pm 250~\si{mm}\) about the shower axis.
To strike a balance between the fidelity of all \gfour{} hits, and computational efficiency,
the energy in each layer is binned into a grid of \(500\times500\) bins (each cell is split into \(5\times5\) bins).
This is a slightly coarser granularity than that of \cctwo{},
and was implemented because lowering the number of points was seen to have minimal impact on the kinematics after projection, while improving the computational efficiency of the model. 
As with \cctwo{}, in each event, this grid is given a uniform random offset of up to half a cell, to reduce discretisation effects. 
Multiple hits may be present in a single bin, and often are near the centre of the shower.
Rather than place all the energy of the hits in the bin in a point at the centre of the bin, a point a is created at the location of the highest energy hit in the bin.
This slightly better reflects the true energy distribution, at minimal cost.
This creates a set of up to \(6,000\) points per shower.
The changes from the previous version, \cctwo{}, are summarised in table~\ref{tab:caloclouds-comparison}.

\begin{table}[h]
    \centering
    \begin{tabular}{l|p{5cm}|p{5cm}}
     & \cctwo{} & \ccthree{} \\
    \hline
    Photon energy range & \(10\) to \(100~\si{GeV}\) & \(1\) to \(127~\si{GeV}\) \\
    \hline
    Photon direction & Directly upwards & Uniformly sampled in upper barrel segment \\
    \hline
    Detector geometry & Real geometry with support structure and staggered cells & Regularised geometry: no gaps in layers for supports  \\
    \hline
    Layer size (in \si{mm}) & $400~\si{mm} \times 400~\si{mm}$ & $500~\si{mm} \times 500~\si{mm}$ \\
    \hline
    Bin granularity (per cell) & $6 \times 6$ bins per cell  & $5 \times 5$ bins per cell  \\
    \hline
    Bin to point conversion & Point placed at centre of bin & Point placed at highest-energy location in bin \\
    \end{tabular}
    \caption{Summary of updates to training data for \ccthree{}}\label{tab:caloclouds-comparison}
\end{table}

\section{\caloc{} Architecture}\label{sec:model}

A wide variety of approaches to fast calorimeter simulation have been attempted.
\caloc{} explores a point based approach.
This has proven particularly advantageous for high granularity calorimeters,
where cell activity is typically sparse and detailed, meaning that grid based approaches would require a prohibitively large number of voxels and parametrised approaches would struggle to replicate the fine structure.

\subsection{Previous \cctwo{} model}\label{sec:cctwo_arch}

A starting point of the existing \cctwo{} model is taken.
A complete description, along with a description of the training data, is available in the existing paper~\cite{Caloclouds2023}, or a somewhat abbreviated description can be found in appendix~\ref{apen:cctwo}.

The key points for the following discussion are that the \caloc{} models are composed of a normalising flow, referred to as \showerf{}, which determines the overall structure of each photon shower, and a diffusion model, which generates each point.
Both models are conditioned to create behaviour appropriate for the incident photon.
The previous \cctwo{} model was only capable of simulating photons at a fixed position, and the model's hyperparameters were adopted from standard practice for these models. 

\subsection{Upgrade methodology}\label{sec:upgradeMethodology}

In this section we describe two elements that significantly shaped choices made in developing \ccthree{}.

\paragraph{Methodology for angular conditioning.} Conditioning on the incident angle is a relatively simple change.
In \cctwo{} the normalising flow was conditioned on \(E_\mathrm{inc}\), the incident particle energy,
in \ccthree{} it will be conditioned on 
$(E_{inc}, \vec{p}_{inc}/|\vec{p}_{inc}|)$ , the incident energy and normalised momentum direction.
Likewise, the diffusion model of \ccthree{} will be conditioned on $(E_{inc}, \vec{p}_{inc}/|\vec{p}_{inc}|)$.
Conditioning the diffusion model on the number of points in the shower, used in \cctwo{}, was found to not have a significant effect and therefore removed.

The decision to use Cartesian coordinates for the direction, rather than a pair of angles, was driven by a desire to avoid degeneracy in the input specifications.
Consider a local polar angle, measuring the deviation of the incident photon from one that travels at \(90\degree\) to the surface of the layers.
When this local polar angle is \(0\) there are an infinite number of azimuthal angles specifying the same direction, which is challenging for interpolation. 
We do observe some minor instability at \((0, 0, 1)\), 
as this point is singular.
Occasionally, this leads to unphysical events from the \showerf{} (with many negative energies or number of points) but it is rare enough (around \(0.1\%\) of the samples) to be handled by rejection sampling.

\paragraph{Methodology for efficiency improvements.} In \cctwo{}, the inference time of the model was dominated by the normalising flow, \showerf{}. 
While the number of parameters in both the normalising flow and the diffusion model have been reduced for \ccthree{}, the focus has been on the flow model.

In order to limit the computational resources (and therefore the carbon footprint) required to investigate variations of normalising flow models, some toy models,
representing expected data characteristics were employed.
These small scale experiments focused on investigating two potential properties of a probability distribution function (pdf); sub-manifolds and first or second order discontinuities. 
Physics data might be expected to contain sub-manifolds, as the number of output parameters generally exceeds the number of input parameters, and would be expected to be approximately smooth, with few first order discontinuities. 
For example, the pdf of the energy in each layer of the detector would be expected to vary smoothly with incident particle energy, and have an approximate sub-manifold defined by adjacent layers having similar energies.
Toy datasets were created from various pdfs, with and without the aforementioned properties.
By observing the training and goodness of fit for normalising flow models with a similar architecture on those toy datasets, we could economically obtain indications of what might be expected in more complex cases.
While this is far from conclusive, it helped guide development of the more expensive training process of the full model.

All expected characteristics of the physics data favoured the use of simple normalising flow models.
By reducing the parameter space of our \showerf{} model, it was seen that the number of parameters could be safely reduced by a factor of 5 (but that a factor of 10 was detrimental to performance).
In a similar fashion, the number of parameters in the diffusion model was reduced by a factor of 4.

\begin{figure}[ht]
    \centering
    \includegraphics[width=0.8\textwidth]{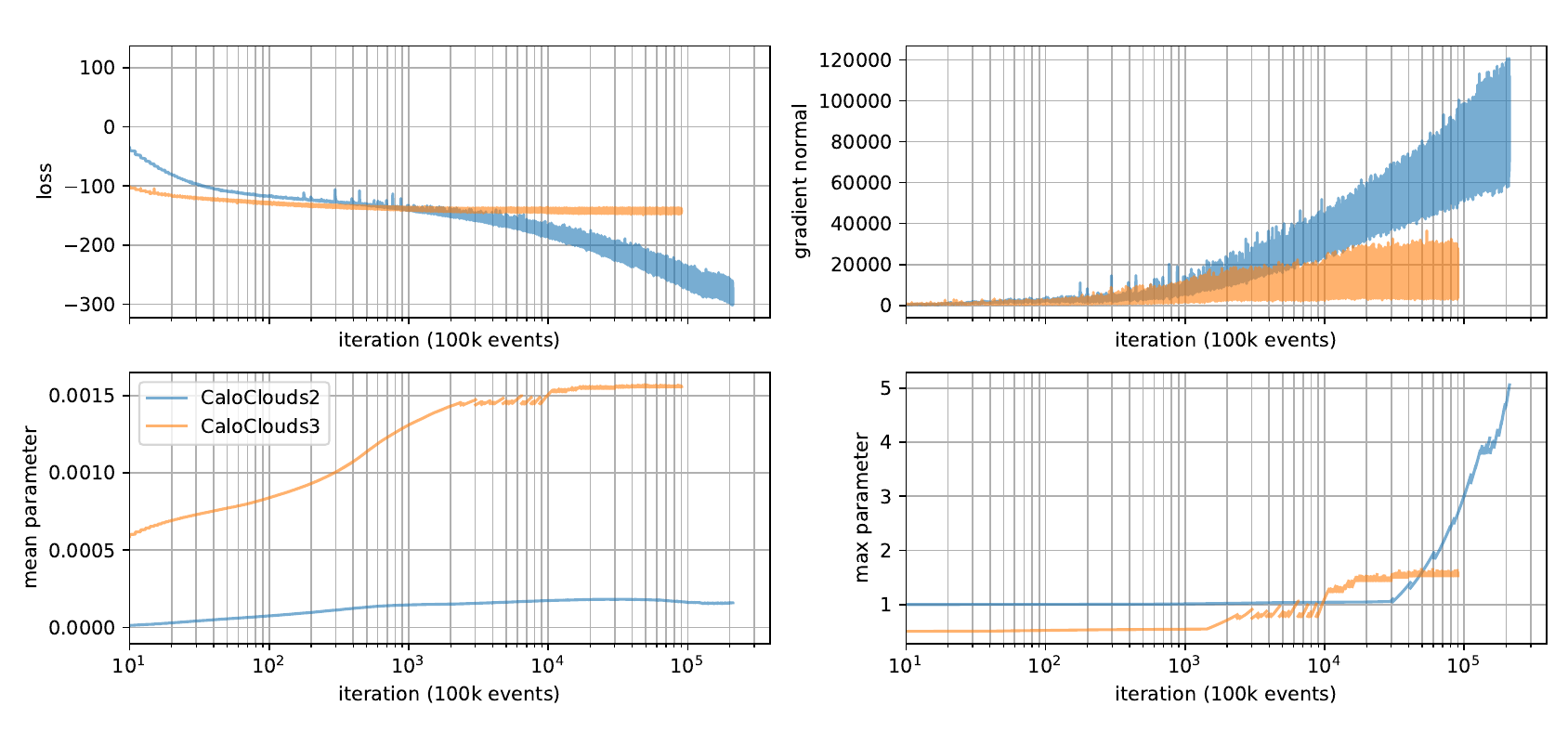}
    \caption{Differences observed while training the \showerf{} model for \cctwo{} and \ccthree{}.
        In the top left, we observe that the training loss of \cctwo{} never plateaus.
        In the top right, we see that the gradient normal of \ccthree{} stabilised during the training, but the
        gradient normal of \cctwo{} continues to increase.
        The relative magnitudes of the mean and maximum parameters in \ccthree{} remain in fairly constant relationship.
    }
    \label{fig:training}
\end{figure}

While the toy models showed no overwhelming failure to learn due to manifold overfitting, sub-manifolds did somewhat impair training stability. 
Identifying and removing sub-manifolds in the normalising flow of our physical model still offered a marginal performance increase.
Comparing the training process of \showerf{} in \cctwo{} and \ccthree{} (figure~\ref{fig:training}), there are indications that a more stable training is taking place in \ccthree{}.

\subsection{\ccthree{} architecture}\label{sec:CC3arch}
This section describes the dataset and finalised design choices for the \ccthree{} model.
It highlights changes made since the previous \cctwo{} model.

\begin{figure}[ht]
    \centering
    \includegraphics[width=0.5\textwidth]{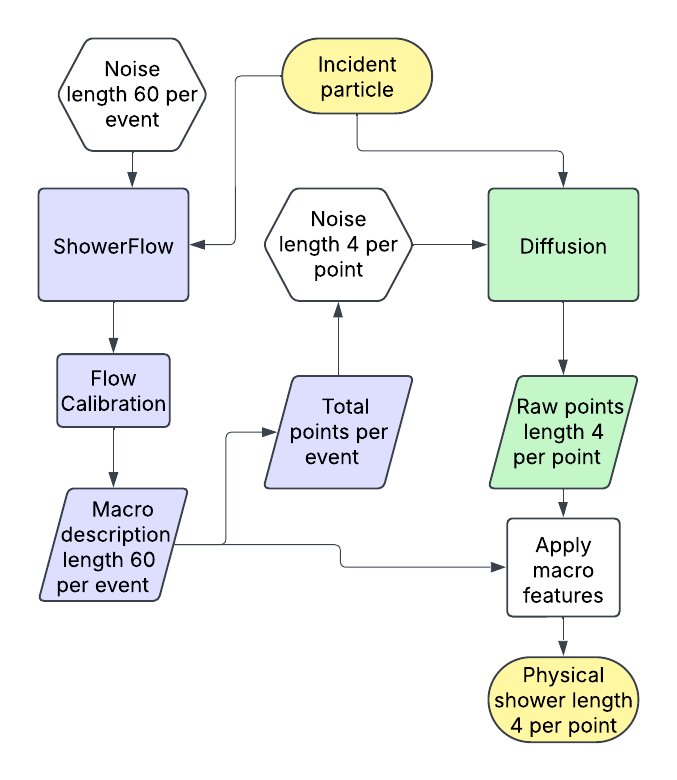}
    \caption{
        The model architecture of \ccthree{}.
        Parts belonging to \showerf{} are in purple, and those belonging to the diffusion model are in green.
        A full description is provided in section~\ref{sec:CC3arch}.
    }
\label{fig:CC3arch}
\end{figure}

As with \cctwo{}, the model uses a normalising flow, referred to as \showerf{}, (shown in purple in figure~\ref{fig:CC3arch}) and a distilled diffusion model (shown in green in figure~\ref{fig:CC3arch}).
\showerf{} is now limited to distributions along the shower axis, specifically the points and energy per layer, and the diffusion model is still used to create individual points.
Both are rescaled by $(E_{inc}, \vec{p}_{inc}/|\vec{p}_{inc}|)$ , the incident energy and normalised momentum direction of the incident photon.
The decision to use Cartesian coordinates (rather than angles) is detailed in section~\ref{sec:upgradeMethodology}.

In \ccthree{}, the information predicted by \showerf{} is
\begin{itemize}
    \item[\(1\) to \(30\)] The number of points in each layer divided by a fixed scaling factor of \(7864\) which rescales the majority of the distribution to be between 0 and 1.
    \item[\(30\) to \(60\)] The visible energy in each layer, divided by a fixed scaling factor of \(3.4\) that rescales the majority of the distribution to be between 0 and 1. 
\end{itemize}

The reductions from the design of \showerf{} in \cctwo{} (see appendix~\ref{apen:cctwo}) remove redundancies that created sub-manifolds in the normalising flow.
The flow interleaves 12 affine couplings and 2 spline couplings.
This has been reduced from the 60 affine couplings and 10 spline couplings used in \cctwo{}, as a much simpler flow was seen to be sufficient to capture the complexity of the distribution,
while being faster at inference time and more stable to train.
A variant of \showerf{} that predicted the log of the energy and points per layer was experimented with, but it was found to be unstable, and so ultimately not pursued.
The Center of Gravity (CoG) of the shower is no longer predicted by \showerf{}, the reasoning for which is discussed later in this section.

The \showerf{} model is trained using an Adam~\cite{adam} optimiser, with a learning rate of \(0.0002\). The batch size is \(62\) and a linear scheduler is used. Due to the high volume of training data, overfitting is not observed.
This output will be reverted to physical values in the step \emph{Apply macro features},
which yields the number of points in each layer, and the energy in each layer. 

Then, like in \cctwo{}, a diffusion model which has been distilled to a single step\footnote{As with \cctwo{}, distillation of the diffusion model is seen to have no adverse impact on the physics performance of the model.}
is used to generate a batch of individual points from an independent and identically distributed (iid) distribution.
Until the generated points are placed in a layer they are referred to as raw points. 
Unlike \cctwo{}, the diffusion model is no longer conditioned on the total number of points.
Removing this conditioning was seen to have no impact on the physics performance of the model;
the conditioning is the same as for \showerf{} (incident particle energy and direction). 
Raw points express their energy on a log scale, such that the energy spectrum of the points covers all real numbers.
These raw points are arranged as a shower during the step \emph{Apply macro features}.
Starting from the first layer the photon reaches 
and working outwards:
\begin{enumerate}
    \item The number of points predicted by \showerf{} for this layer are selected from the raw points generated from the diffusion model, starting with those generated at the base of the shower.
    These points are moved along the shower axis to sit at the centre of the layer and assigned to that layer, removing them from the raw points.
    \item The energy of the selected points is returned to a linear scale.
    \item The total energy of this layer is rescaled to match the energy predicted by \showerf{}. 
\end{enumerate}

This yields a shower in physical units, which can be projected back into detector cells.
Two fixed scale factors are applied to correct for bias in occupancy and cell energy.
This is simplified from \cctwo{}, where occupancy was corrected with a polynomial fit.

\paragraph{Projection to detector geometry.} On a basic level, detector readout can be mimicked by converting the energy deposits back to detector coordinates and allocating them to cells using records of the cell and layer boundaries.
To calculate the energy recorded by a cell, the energy of all the points that fall within each cell are summed. Energy that falls into gaps between sensitive cells is abandoned. 
This is the method used to emulate detector readout with which \cctwo{} and \ccthree{} are compared.

When working with only \ccthree{} a more extensive framework can also facilitate reconstruction. The \ddforhep{}~\cite{DD4hep} toolkit exists to provide detector descriptions.
Built on top of this, the \ddsim{} and \ddml{}~\cite{key4hep_DDML} libraries collectively run simulations.
A full \gfour{} simulation can be run, or an augmented \gfour{} simulation, in which particles identified as suitable for a fast simulation model will be delegated to that fast simulation. 
This is only possible with \ccthree{} as it requires angular conditioning, which \cctwo{} lacks.

To use a fast simulation model in this manner, it must be compiled. This can be achieved using \torchscript{}~\cite{torchscript}, which compiles the models in a language designed to be cross compatible with C++. Once the model has been compiled, triggers are defined using \ddsim{}'s python interface, to replace photons in the full simulation that \ccthree{} is able to simulate.

Once simulation, using \gfour{} or \gfour{} and \ccthree{}, is complete, the energy deposits created are projected back into the detector geometry supplied by \ddforhep{} to mimic real detector readout, and reconstruction tools are applied.

\paragraph{Removal of CoG calibration.} 
For a perfect detector, conservation of momentum would ensure that the position weighted energy sum of the shower's energy depositions was exactly on the incident particle axis, also refered to as the shower axis, \(\hat{s}\).
The ILD has a sampling fraction of around \(1\%\) in the ECAL, therefore missing energy may break this symmetry.

In this paper, we define CoG as the weighted sum of the shower's energy, in a coordinate system relative to the shower axis.
Let \(c\) be the vector from the point the photon meets the calorimeter to the energy weighted mean of the shower, and \(\hat{z}\) be the detector global \(z\) direction, that is the direction of the beam then the centre of gravity can be expressed as
\begin{equation*}
\mathrm{center~of~gravity~X} = ((\hat{s}\times\hat{z})\times\hat{s})\cdot c
\end{equation*}
and
\begin{equation*}
\mathrm{center~of~gravity~Y} = (\hat{s}\times\hat{z})\cdot c.
\end{equation*}
Typically, both these values are small, because an even sampling of the random energy deposits means the CoG remains close to the shower axis. 
When the CoG does deviate significantly from the shower axis,
it has been shifted by high energy secondary particles scattered at wide angles.
As the diffusion model draws points iid, these correlated structures cannot be replicated, and so the tails of the 
CoG distribution are not well modelled by the diffusion model. 

Previously, in \cctwo{}, \showerf{} predicted values for the CoG perpendicular to the shower axis, and all points in the shower were shifted to match in calibration.
This did achieve a good CoG distribution, 
however it is unphysical because the tails of the \gfour{} CoG distribution 
are not formed by shifting the bulk of the points, but
by correlated, off centre, clusters of hits.

Given this objection, it was decided that 
the \(x\) and \(y\) centre of gravity that was naturally yielded by the diffusion model would be left as is.
It will be seen in the results section that this only marginally degrades the centre of gravity profile in \(x\) and \(y\).

We also consider a third axis for the CoG, in the direction perpendicular to the layers of the detector. This is referred to as center of gravity Z, and is determined by the energy per layer, as modelled by \showerf{}. 


\section{Physics Accuracy}\label{sec:accuracy}

To evaluate the model, first a number of kinematics are compared on the unreconstructed detector output.
Having significantly altered the model to allow angular conditioning and improve speed,
we compare kinematics between \cctwo{} and \ccthree{} to verify that the performance has not been impaired. 
Then the performance of \ccthree{} is investigated at fixed energies, as varying the incident energy of the photon has a significant impact on the shower behaviour.
Finally, the replication of shower angle is investigated.

\subsection{Impact of the changes from \cctwo{} to \ccthree{}}

\begin{figure}[ht]
    \centering
\includegraphics[width=0.9\textwidth]{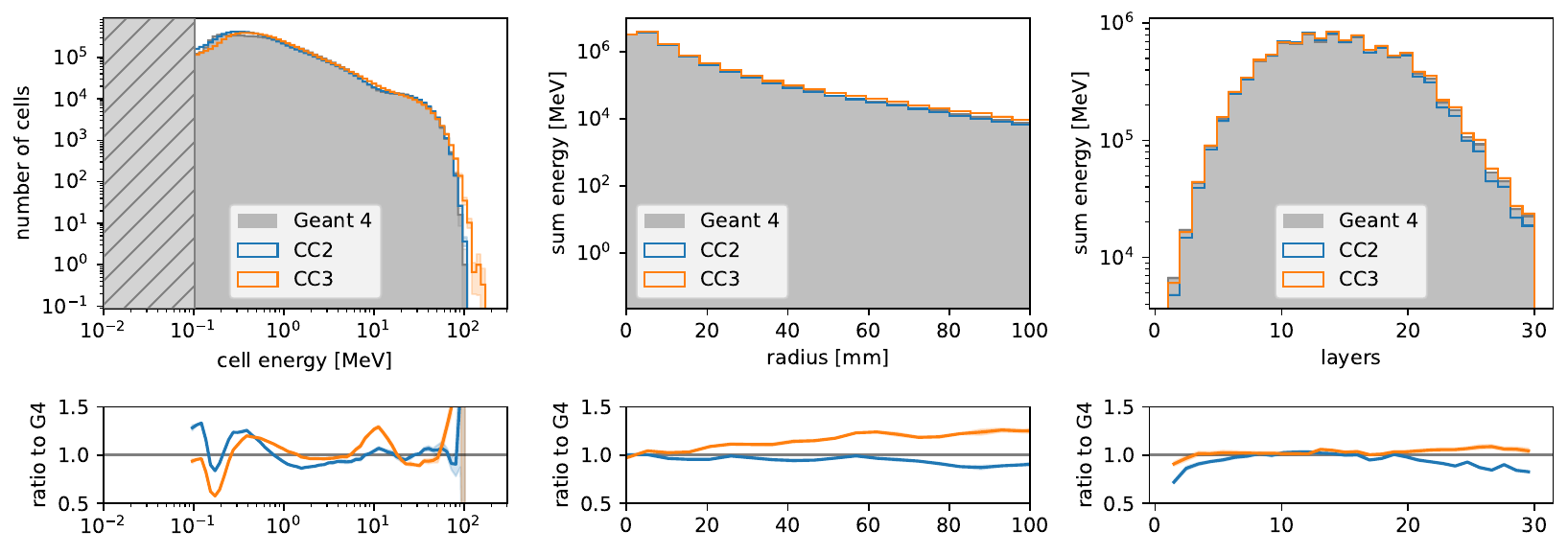}
    \caption{
        Energy distributions of \cctwo{} (CC2) and \ccthree{} (CC3), in comparison to \gfour{}.
        Top row are the distributions themselves, averaged over three different seeds,
        with error bars from the standard deviation.
        Bottom row are ratio plots between each model and \gfour{}, with errors propagated from
        the distributions.
        Left, histogram of the energy of each cell, with the MIP cut marked as a hatched box.
        Centre, radial profile of the energy.
        Right, distribution of the energy through the layers.
    }
\label{fig:3key_0}
\end{figure}

\begin{figure}[ht]
    \centering
\includegraphics[width=0.9\textwidth]{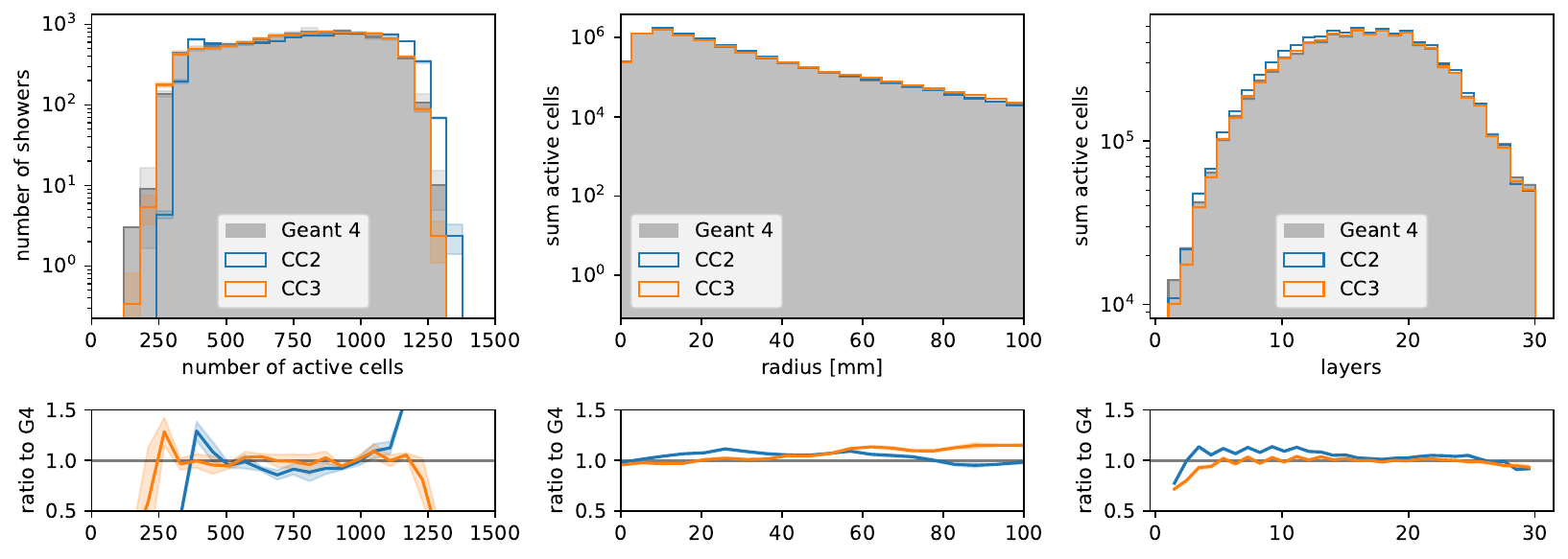}
    \caption{
        Occupancy distributions of \cctwo{} (CC2) and \ccthree{} (CC3), in comparison to \gfour{}.
        Top row are the distributions themselves, averaged over three different seeds,
        with error bars defined by the standard deviation.
        Bottom row are ratio plots between each model and \gfour{}, with errors propagated from
        the distributions.
        Left, histogram of the occupancy of the calorimeter in each shower.
        Centre, the number of active cells in concentric rings about the shower axis.
        Right, the number of active cells in each layer.
    }
\label{fig:3key_1}
\end{figure}

\begin{figure}[ht]
    \centering
\includegraphics[width=0.9\textwidth]{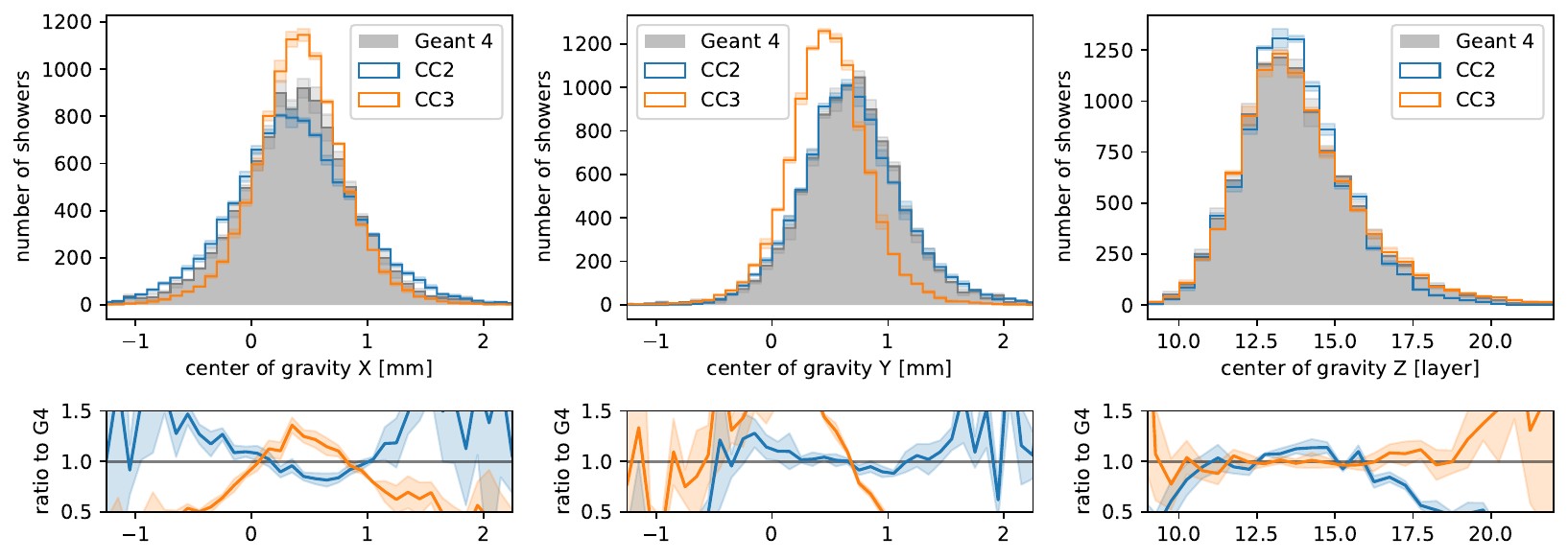}
    \caption{
        Centre of Gravity for showers from \cctwo{} (CC2) and \ccthree{} (CC3), in comparison to \gfour{}.
        Top row are the distributions themselves, averaged over three different seeds,
        with error bars defined by the standard deviation.
        Bottom row are ratio plots between each model and \gfour{}, with errors propagated from
        the distributions.
        Left and centre are the centre of gravity in \(x\) and \(y\) respectively,
        the directions perpendicular to the shower axis.
        Right is the centre of gravity parallel to the shower axis,
        expressed as a position in the 30 layers.
    }
\label{fig:cog}
\end{figure}

The fidelity of photons with orthogonal impact, between \(10\) and \(90~\si{GeV}\), is compared between \cctwo{}, \ccthree{} and \gfour{},
since this is the only region in which \cctwo{} can produce output.
The results are projected into a realistic detector map, as described at the end of section~\ref{sec:CC3arch},
and key shower observables calculated -- these observables are shown in figures~\ref{fig:3key_0},~\ref{fig:3key_1} and~\ref{fig:cog}.

In the comparison, \cctwo{} is being trained on photons generated the same way as those used as benchmark,
including all detector geometry effects, while the training dataset of \ccthree{} is as described in section~\ref{sec:preprocessing}.

This does give \cctwo{} an advantage in detector geometry, but this is unavoidable as \cctwo{}
requires orthogonal photons for training.
Barring one small change, it is trained in the same was as it was in its debut paper~\cite{CaloClouds2}. 
That is, photons with orthogonal impact, between \(10\) and \(90~\si{GeV}\),
with slightly higher granularity and a realistic detector, including all supporting structure. 
However, when generating the previous dataset for training \cctwo{},
the photon gun was placed just inside the calorimeter, before the first sensitive layer,
but after the first absorber.
This was an error, and it caused the first sensitive layer to be overpopulated,
so the dataset for this work has been regenerated with the gun pulled back,
to sit properly outside the calorimeter.

Figure~\ref{fig:3key_0} focuses on the energy profiles.
The leftmost plot is a histogram with the energy of each cell,
the far left 
of which is blocked off to represent the MIP (minimally ionising particle) cut.
Cells with less than than half the energy of a MIP are conventionally removed to reduce issues of electronic noise.
Just to the right of this, at \(\approx0.2~\si{MeV}\), the MIP peak can be seen.
While neither model quite replicates the MIP peak, \cctwo{} is a little closer.
On the right hand of this first plot,
 \ccthree{} is seen to slightly overstate the high energy tail of the cell energies.
Overall, this plot is perhaps most impacted by the reduction in granularity of the training data,
however the effect is still marginal.

The central plot of figure~\ref{fig:3key_0} shows the radial profile of the energy.
Detector cells are assigned to a bin of this histogram according to their radial distance from the shower axis, and the energy from that cell added to the bin total.
One can observe a slight dip at 0 radius because the impact point is in the centre of a cell, so the first bin is only half a cell wide. 
Overall, both models replicate the \gfour{} data well, with \cctwo{} producing a marginally flatter ratio, and \ccthree{} tending to overestimate the amount of energy in the tails.
We do not know why \ccthree{} overestimates the tails of the radial distribution.

The right hand plot of figure~\ref{fig:3key_0} shows the energy per layer.
This plot shows an alternating pattern, 
as the silicon sensor layers are constructed in pairs, and mounted back to back on a tungsten absorber layer. 
\ccthree{} replicates the distribution the best, achieving a remarkably flat ratio throughout,
although both models perform well.

Figure~\ref{fig:3key_1} focuses on the occupancy profiles.
The leftmost plot is a histogram with the occupancy of the calorimeter in each shower.
As we simulate a continuous range of energies (\(10~\si{GeV}\) to \(90~\si{GeV}\)), this
distribution has a relatively flat profile.
Both models are seen to perform well, but \cctwo{} has a slight tendency to overestimate the occupancy.

In the centre plot of figure~\ref{fig:3key_1}, the number of active cells per unit radius is shown.
Like the radial energy, detector cells are assigned to a bin of this histogram according to their radial distance from the shower axis, and the number of cells with energy above half a MIP is counted towards the bin total.
Towards \(0\) radius, bins encompass a smaller volume of the detector, creating a dip in the number of active cells.
In this metric, both models are excellent, with \ccthree{} recreating the centre of the distribution the best.

The rightmost of the three plots shows the number of active cells per layer (or the occupancy in each layer).
The ability of the models to replicate this pattern is defined by the \showerf{} component,
which specifies the points in each layer.
Due to the alternating absorber placement, every other layer of the calorimeter tends to accumulate more hits.
This can be seen in figure~\ref{fig:3key_0} as an excess of energy in every other layer, and in figure~\ref{fig:3key_1} as an excess of hits in every other layer.
The change in absorber thickness at layer \(20\) is also somewhat visible here.
Both models replicate these effects well, but \ccthree{} has a flatter ratio to \gfour{}.

Figure~\ref{fig:cog} shows the centre of gravity of the two models against the centre of gravity produced by \gfour{}.
Here, we expect worse performance in the perpendicular direction from \ccthree{},
since it has had the centre of gravity conditioning from \showerf{} removed,
as per the reasoning given in section~\ref{sec:CC3arch},
and additionally, it does not have the advantage of training on the specific detector location with supporting structure.
The first two plots, left and centre, show the perpendicular directions. As expected, 
\cctwo{} matches these better, with \ccthree{} slightly underestimating the width of the distribution, and not quite replicating the offset.

On the right, the centre of gravity along the shower axis is shown, expressed as a location in the 30 layers.
In this \ccthree{} provides a marginally better fit, but both models perform acceptably.

\begin{table}[!ht]
\centering

\begin{tabular}{l|cc}
\multicolumn{1}{c|}{Variable}  & \multicolumn{2}{c}{Jenson Shannon divergence} \\
\multicolumn{1}{c|}{compared} & \cctwo{} & \ccthree{} \\
\hline
total visible energy & $\mathbf{1.52(52) \times 10^{-3}}$ & $3.75(77) \times 10^{-3}$ \\
energy per cell & $\mathbf{2.42(4) \times 10^{-3}}$ & $4.54(6) \times 10^{-3}$ \\
energy per layer & $2.17(17) \times 10^{-4}$ & $\mathbf{4.01(96) \times 10^{-5}}$ \\
energy per unit radius & $\mathbf{0.64(16) \times 10^{-4}}$ & $3.15(35) \times 10^{-4}$ \\
total occupancy & $1.42(14) \times 10^{-2}$ & $\mathbf{0.59(29) \times 10^{-3}}$ \\
occupancy per layer & $2.12(12) \times 10^{-4}$ & $\mathbf{9.55(93) \times 10^{-5}}$ \\
occupancy per unit radius & $\mathbf{1.33(12) \times 10^{-4}}$ & $2.91(13) \times 10^{-4}$ \\
center of gravity in X & $\mathbf{4.38(79) \times 10^{-3}}$ & $0.96(11) \times 10^{-2}$ \\
center of gravity in Y & $\mathbf{2.20(48) \times 10^{-3}}$ & $3.05(19) \times 10^{-2}$ \\
center of gravity in Z & $5.00(75) \times 10^{-3}$ & $\mathbf{0.90(31) \times 10^{-3}}$ \\
\end{tabular}
\caption{All metrics in this sub-section are summarised by calculating a Jenson Shannon divergence between \gfour{} and each model.
The distance is calculated on histograms that combine all energies used, shown in appendix section~\ref{appendix3},
as figures~\ref{fig:3key_allang_0} and~\ref{fig:3key_allang_1}.
An epsilon of \(10^{-10}\) is added to empty bins to prevent division by zero.
Errors have been propagated from the histograms.
The value in bold is the better of the two scores.}\label{tab:cc23_metrics} 
\end{table}

Since the changes from \cctwo{} to \ccthree{} are significant, involving both a reduction in size,
and the extension of the model to cover incident particles of all angles,
it is satisfactory that the metrics in table~\ref{tab:cc23_metrics} indicate comparable global performance.
The changes have improved some fits, and degraded others, but for the most part \ccthree{} remains at the same order of magnitude or better than \cctwo{}.
Only for the centre of gravity in X/Y does \cctwo{} perform a lot better, and this is due to the removal of CoG calibration, which is discussed in section~\ref{sec:CC3arch}.
Overall it appears that the upgrades to \ccthree{} have not adversely impacted the performance of the model.

\subsection{Performance of \ccthree{} at random angles and fixed energies}\label{sec:fixed_energies}

Having established that \ccthree{} can replicate photons travelling along the \(z\) axis,
the next assessment looks at photons all across a barrel segment to evaluate the behaviour of \ccthree{} across all angles. The following observables are computed at the reconstruction level, after the model has been integrated into the simulation pipeline.

We now fix the energy of the incident particle in increments of \(10~\si{GeV}\) from \(10~\si{GeV}\) to \(100~\si{GeV}\).
For legibility the histograms only present in full \(10~\si{GeV}\), \(50~\si{GeV}\) and \(100~\si{GeV}\). 
All other energies are included in summary statistics plotted in figure~\ref{fig:ang_js_spec}
and plotted collectively in figure~\ref{fig:3key_allang_1}.

\begin{figure}[ht]
    \centering
    \includegraphics[width=0.9\textwidth]{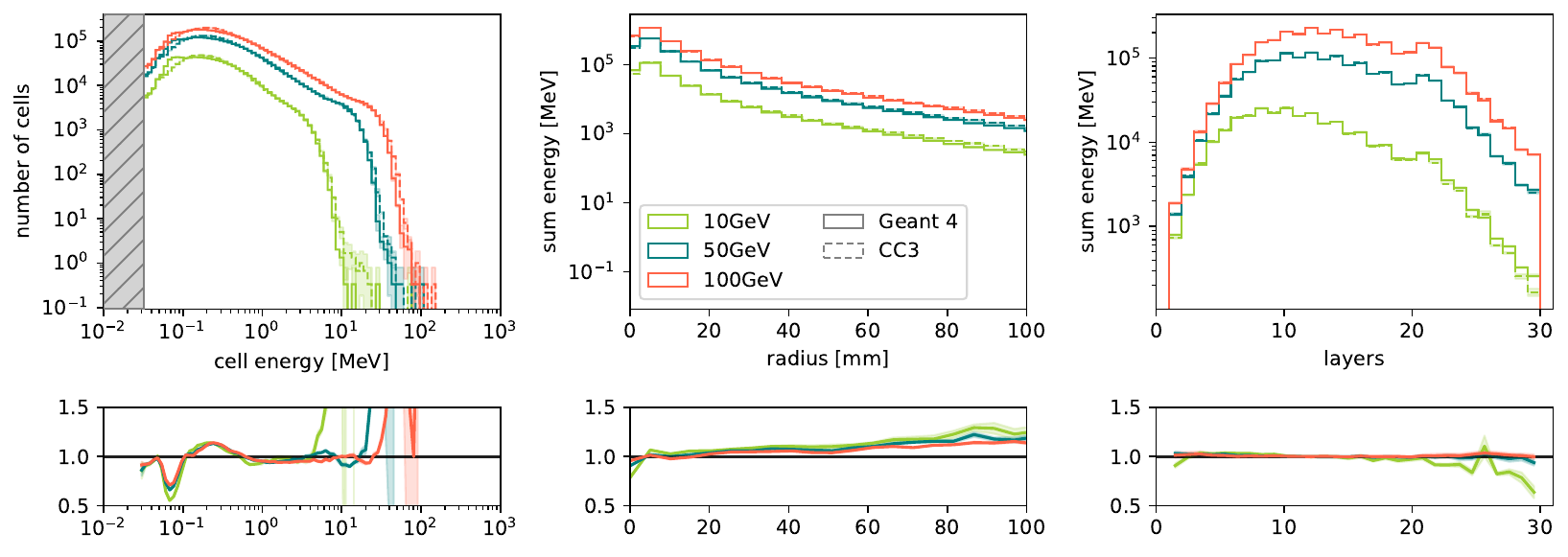}
    \caption{
        Energy spectrum of \ccthree{} (CC3) at random angles and fixed incident energies.
        For legibility, the plot only shows three of the ten photon incident energies tested;
        \(10~\si{GeV}\), \(50~\si{GeV}\) and \(100~\si{GeV}\).
        In the upper plot, \gfour{} is shown as a solid line, and \ccthree{} as a dashed line,
        though for some distributions the lines are too close to distinguish.
        The lower plot shows ratios between \gfour{} and \ccthree{}.
        Layout is otherwise the same as in figure~\ref{fig:3key_0}.
    }\label{fig:3key_ang_0}
\end{figure}

Figure~\ref{fig:3key_ang_0} shows examples of the different energy distributions. 
The leftmost plot shows the per-cell energy distribution, with the far left blocked off by the MIP cut applied in reconstruction.
For the majority of the spectrum, all three energies produce a flat ratio, only
somewhat overestimating the tail of high energy cells.
The central plot of figure~\ref{fig:3key_ang_0} shows the radial profile of the energy.
Detector cells are assigned to a bin of this histogram according to their radial distance from the shower axis, and the energy from that cell added to the bin total.
In the right hand plot the energy per layer is seen.
At all energies, these distributions are well replicated, but
the ratio is seen to improve somewhat at higher incident particle energies.

\begin{figure}[ht]
    \centering
    \includegraphics[width=0.9\textwidth]{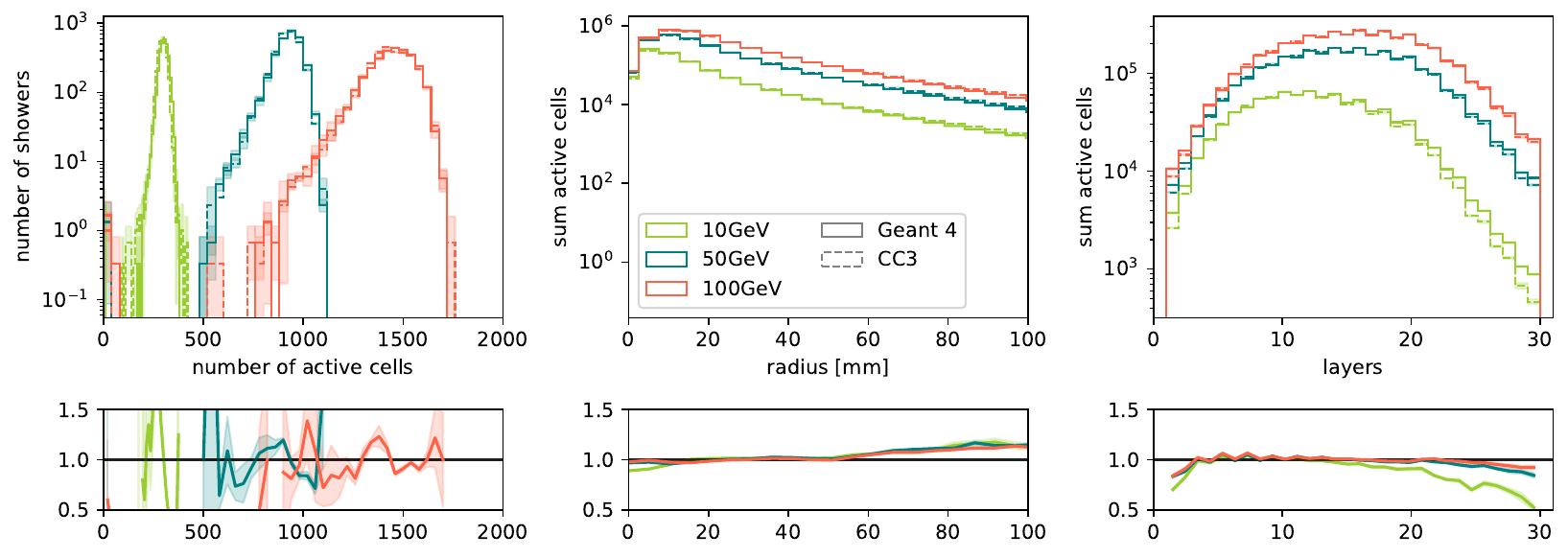}
    \caption{
        Occupancy spectrum of \ccthree{} (CC3) at random angles and fixed energies.
        For legibility, the plot only shows three of the ten photon incident energies tested;
        \(10~\si{GeV}\), \(50~\si{GeV}\) and \(100~\si{GeV}\).
        In the upper plot, \gfour{} is shown as a solid line, and \ccthree{} as a dashed line,
        though for some distributions the lines are too close to distinguish.
        The lower plot shows ratios between \gfour{} and \ccthree{}.
        Layout is otherwise the same as in figure~\ref{fig:3key_1}.
    }\label{fig:3key_ang_1}
\end{figure}

Figure~\ref{fig:3key_ang_1}, shows examples of the occupancy spectrum of \ccthree{} at random angles and fixed energies.
The leftmost plot is a histogram of total occupancy in each shower,
while the higher two energies show reasonable agreement, it's clear that the model does not perform as well at the lowest energy.

The central plot of figure~\ref{fig:3key_ang_1} shows the radial profile of the occupancy.
Like the radial energy, detector cells are assigned to a bin of this histogram according to their radial distance from the shower axis, and the number of cells with energy above half a MIP is counted towards the bin total.
Towards \(0\) radius, bins encompass a smaller volume of the detector, creating a dip in the number of active cells.
In the right hand plot the occupancy per layer can be seen.
At all energies, these distributions are well replicated, with a more pronounced
deviation at \(10~\si{GeV}\) in the occupancy per layer.

\begin{figure}[ht]
    \centering
    \includegraphics[width=0.8\textwidth]{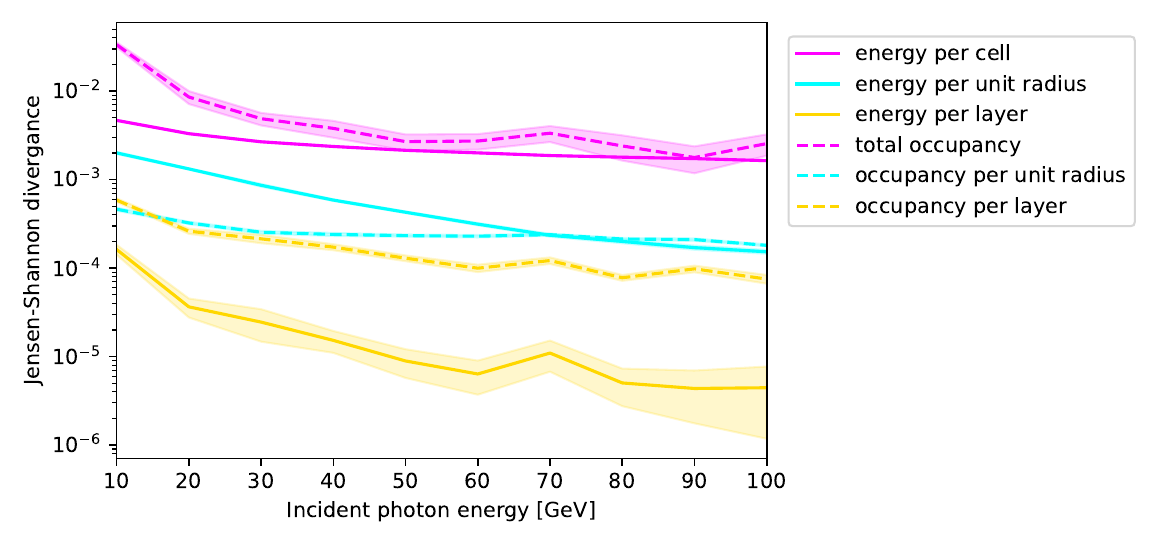}
    \caption{
        Kinematic performance of \ccthree{} at random angles and fixed energies.
        The Jenson Shannon divergence between \gfour{} and \ccthree{} for the energy spectrums
        (see figure~\ref{fig:3key_ang_0}) at each incident energy 
        and the occupancy spectrums (see figure~\ref{fig:3key_ang_1})
        are plotted against the incident photon energy.
    }\label{fig:ang_js_spec}
\end{figure}

\begin{table}[!ht]
\resizebox{\columnwidth}{!}{
\centering
\begin{tabular}{c|cccccc}
\makecell{} & \makecell{energy per \\ cell} & \makecell{energy per \\ layer} & \makecell{energy per \\ unit radius} & \makecell{occupancy \\ per layer} & \makecell{occupancy \\ per unit \\ radius} & \makecell{total \\ occupancy} \\
\hline
\ccthree{} & \(2.06(1) \times 10^{-3}\) & \(0.50(11) \times 10^{-5}\) & \(2.95(4) \times 10^{-4}\) & \(1.13(4) \times 10^{-4}\) & \(2.18(2) \times 10^{-4}\) & \(3.62(26) \times 10^{-3}\) \\
\end{tabular}
}
\caption{All of the plots in this sub-section are summarised by calculating a Jenson Shannon distance between \gfour{} and each model.
An epsilon of \(10^{-10}\) is added to empty bins to prevent division by zero.
Errors have been propagated from the histograms.
}\label{tab:fixed_energies_metrics}
\end{table}

For these comparisons we make two forms of summary statistic.
The first is the Jenson Shannon distance between \gfour{} and \ccthree{} at each energy separately.
This is plotted against the incident energy in figure~\ref{fig:ang_js_spec}.
In all cases, the clear conclusion is that \ccthree{} performs best at higher energies.
The main reason for this is that the diffusion model makes an iid assumption about the points it is creating,
and this iid assumption is closest to the truth at higher energies.
In figure~\ref{fig:ExampleShower}, examples of showers produced by \gfour{} at \(10~\si{GeV}\), \(50~\si{GeV}\) and \(100~\si{GeV}\) are given.
Examining these, more structure can be seen at \(10~\si{GeV}\), but by \(100~\si{GeV}\) this structure is obscured by the number of hits.
As \ccthree{} is incapable of replicating tracks and other internal hit correlations, it performs best at higher energies where such things have been
washed out.

In addition, in table~\ref{tab:fixed_energies_metrics}, we calculate a Jenson Shannon distance between the histograms
of all energy increments combined.
The histograms themselves can be found in the appendix section~\ref{appendix3}.

Overall, the kinematics seem well matched in this dataset representing all angles.

\subsection{Angular reconstruction}\label{sec:angular_reconstruction}

One further point that must be investigated is the replication of the internal angle of the shower.
The internal angle of a particle shower is defined by the direction that would be reconstructed for that shower's incident particle,
without any assumptions on the origin of the incident particle.
This is distinct from the angle of the incident particle that would be reconstructed with the assumption
that the incident particle originated from the interaction point at the centre of the detector.
For searches for long lived particles, such as a dark matter particle, the reconstruction of the internal angle is particularly important.

The dataset used for this comparison is the same as in section~\ref{sec:fixed_energies}, but all ten energies are used in combination.

A common method for calculating the internal angle of the shower is to take the first component of an energy weighted PCA (principle component analysis) applied to
all the reconstructed hits in the shower.
When calculating the internal angle by PCA of 
hits, we observed systematic biases in the angle calculated in this way.
This bias is caused by a combination of factors.
Firstly, the recorded shower misses any secondary particles that were emitted before the first layer of the detector.
When the shower is at an angle, these missed particles are no longer symmetrically distributed about the shower axis.
Secondly, secondary particles travelling at steep and shallow angles experience different path lengths in the detector. 
When the shower axis is at an angle, the secondary particles from one side of the axis experience a different path length to those emitted on the other side, causing an imbalance in their detection.
Further discussion, and illustration of these biases can be found in the appendix section~\ref{appendix2}.

The bias can be quantified as an angular error;
\[\textrm{angular error} = \cos^{-1}\left(\frac{\vec{p}_{\mathrm{PCA}} \cdot \vec{p}_\mathrm{inc}}{|\vec{p}_{\mathrm{PCA}}||\vec{p}_\mathrm{inc}|}\right).\]
Where \(\vec{p}_\mathrm{PCA}\) is the direction of the photon reconstructed by PCA, and \(\vec{p}_\mathrm{inc}\) is the true direction of the incident photon.
Or alternatively, broken down into a \(\phi\) component that rotates in the plane perpendicular to the beam axis and a \(\theta\) component that measures the angle to the beam axis;
\begin{align*}
  \textrm{error in }\phi = \tan^{-1}\left(\frac{\vec{p}_{\textrm{PCA}; x}}{\vec{p}_{\textrm{PCA}; y}}\right) - \tan^{-1}\left(\frac{\vec{p}_{\textrm{inc}; x}}{\vec{p}_{\textrm{inc}; y}}\right), &&
  \textrm{error in }\theta = \cos^{-1}\left(\frac{\vec{p}_{\textrm{PCA}; z}}{|\vec{p}_\textrm{PCA}|}\right) - \cos^{-1}\left(\frac{\vec{p}_{\textrm{inc}; z}}{|\vec{p}_\textrm{inc}|}\right).
\end{align*}

\begin{figure}[ht]
    \centering
    \includegraphics[width=0.6\textwidth]{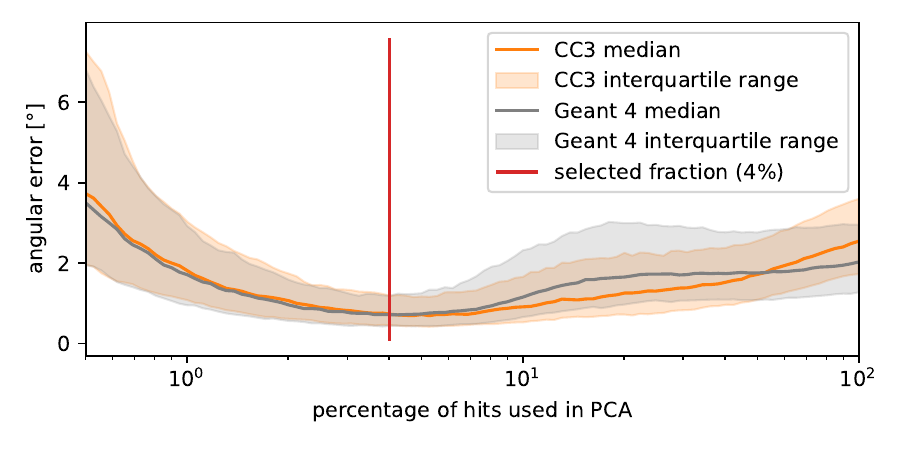}
    \caption{
        Error of the internal angle as a percentage of highest energy hits used in the energy weighted PCA calculation.
        The angular error is the angular distance between the direction of the first PCA component,
        and the true direction of the incident particle.
    }\label{fig:angular_error_scan}
\end{figure}

As the bias on this error is due to off-axis secondary hits, the simplest way to remove it is to do a PCA with only the highest energy hits in the shower.
To rephrase this, rather than calculating the PCA on all hits in the shower, a limited percentage of the highest energy hits are selected,
and a PCA is calculated on this subset.
A scan over the percentage used is shown in figure~\ref{fig:angular_error_scan}.
It is seen that not only does the error itself decrease, but the variance of the error is also reduced.
With this we justify our preference for calculating the internal angle of the shower using the highest \(4\%\) of hit energies in the shower.
Metrics calculated with only this highest energy fraction are referred to as improved metrics.

\begin{figure}[ht]
    \centering
    \includegraphics[width=0.9\textwidth]{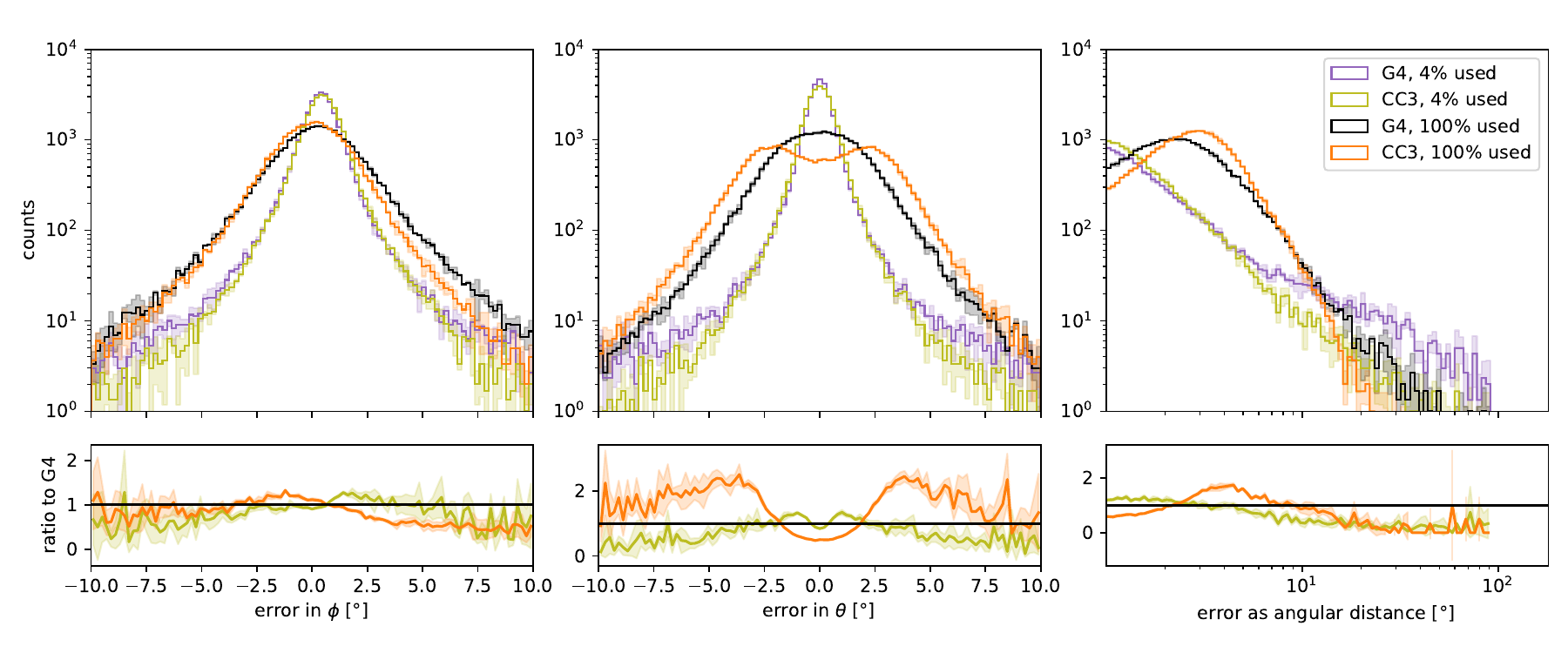}
    \caption{
        Angular errors of \ccthree{} compared to \gfour{}.
        The left hand plot compares error in the azimuthal direction about the detector barrel,
        and the central plot compares error in the polar direction about the detector barrel (angle to the beam).
        The right hand plot compares the angular distance between the reconstructed internal angle and the true direction of the incident particle.
        Each plot, shows showers from two models, and two calculations are given, one using all of the hits, and then the improved metric, using only the highest energy \(4\%\) of the hits.
    }
\label{fig:angular_error_hist}
\end{figure}

Figure~\ref{fig:angular_error_hist} shows the error on the internal angle of \ccthree{} and \gfour{},
considering both the calculation utilising \(100\%\) of the hits, and the highest \(4\%\) of hit energies.
The error in both directions, and in the absolute angle, is considerably improved by taking the highest \(4\%\) of hit energies.

When using only the highest \(4\%\) of hit energies, \ccthree{} produces an internal angle spectrum that is a much better match for \gfour{}.
This would indicate that while the tilt of the highest energy hits is well replicated, the lower energy hits do not have such a good angular placement in \ccthree{}.
Particularly striking, is the double peak in the \(\theta\) angular error of \ccthree{} when calculating with \(100\%\) of the hits.
The double peak is caused by backscattered hits tilting the shower towards the beam direction.
Plots that better illustrate this concept can be seen in the appendix section~\ref{appendix2}.
The difference in the \(\theta\) error spectrum specifically indicates that the backscattering hits are incorrectly placed by \ccthree{}.

Despite the issues with modelling the angle of the lower energy hits, the angular distance distributions of \ccthree{} and \gfour{} remain within a factor of \(2\) when calculating the internal angle using the classical PCA on \(100\%\) of the hits as can be seen in the ratio.
When using the improved metric for internal angle, the agreement with \gfour{} becomes good for all angular metrics, including a proper replication of the \(\theta\) shape.
Summary statistics are shown in table~\ref{tab:angular_errors};
every single metric improves at \(4\%\).
Due to the shape of the barrel segments, the sample spans a smaller range in \(\phi\), compared to \(\theta\), see section~\ref{sec:preprocessing},
therefore it is to be expected that the \(\theta\) and overall angular error metrics will improve more than \(\phi\).
As the internal angle is a key feature in searches of long lived particles,
replication of this metric provides additional use cases.

\begin{table}[!ht]
\centering
\begin{tabular}{c|ccc}
    \makecell{} & \makecell{\(\phi\)} & \makecell{\(\theta\)} & \makecell{angular error} \\
\hline
4\% used  & \(\mathbf{1.96(21) \times 10^{-3}}\) & \(\mathbf{3.76(26) \times 10^{-3}}\) & \(\mathbf{7.75(65) \times 10^{-3}}\)\\
100\% used  & \(5.02(33) \times 10^{-3}\) & \(3.80(8) \times 10^{-2}\) & \(1.49(7) \times 10^{-2}\)\\
\end{tabular}
\caption{
    Jenson Shannon distances between the angular error spectrum as seen in figures~\ref{fig:angular_error_hist} of \ccthree{} and \gfour{}.
}\label{tab:angular_errors}
\end{table}

\section{Di-Photon Separation}\label{sec:reco}

While replicating the kinematics is promising, it doesn't address the true purpose of a fast model, which is to
replace the full simulation in a reconstruction.
The only reliable way to evaluate a fast simulation model is to compare the output of a reconstruction 
applied to showers simulated with the fast model to the output of a reconstruction applied to \gfour{} samples.

Reconstruction performance on a wider variety of metrics, across a range of fast simulation models,
is presented in a companion paper to this one, a first full physics benchmark for highly granular calorimeter surrogates~\cite{benchmarks}.
To validate this model with a basic example, we show the performance of \ccthree{} on di-photon separation.

\begin{figure}[ht]
    \begin{minipage}[c]{0.4\textwidth}
    \includegraphics[width=\textwidth]{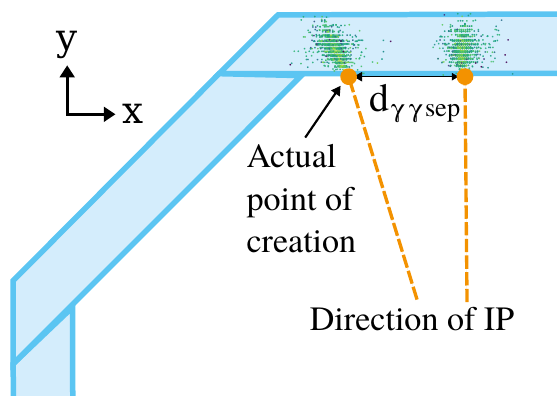}
    \end{minipage}\hfill
    \begin{minipage}[c]{0.57\textwidth}
    \caption{
    Schematic, describing the setup used to test di-photon separation.
    Two photons are created at random points (depicted as orange dots) at the surface of the ECAL (depicted in blue).
    Their direction is consistant with photons originating from the IP.
    The distance, \(d_{\gamma\gamma\mathrm{sep}}\), is measured at the surface of the calorimeter, where the photons are created.
    }\label{fig:diphot_scheme}
    \end{minipage}
\end{figure}

The di-photon separation is a measure of how far apart two photon showers must be to both be resolved in reconstruction.
It is computed by firing a pair of photons with the same energy
at random distances in close proximity
into a barrel segment,
then using standard reconstruction software in \keyforhep{}~\cite{key4hep}
the event is reconstructed and the number of photons are counted.
A schematic of this, defining the distance used, can be seen in figure~\ref{fig:diphot_scheme}.

By basic physical constraints, two photons with exactly the same direction cannot be distinguished from one high energy photon.
As the separation between a pair of photons grows, the likelihood of reconstructing both photons grows,
until the likelihood of reconstructing two photons becomes the product of the likelihood of reconstructing each individual photon.
The progression from at most a single reconstructed photon to two reconstructed photons depends strongly on the radial profile of the shower, and will only match \gfour{} if it is described correctly.

\begin{figure}[ht]
    \centering
    \includegraphics[width=0.3\textwidth]{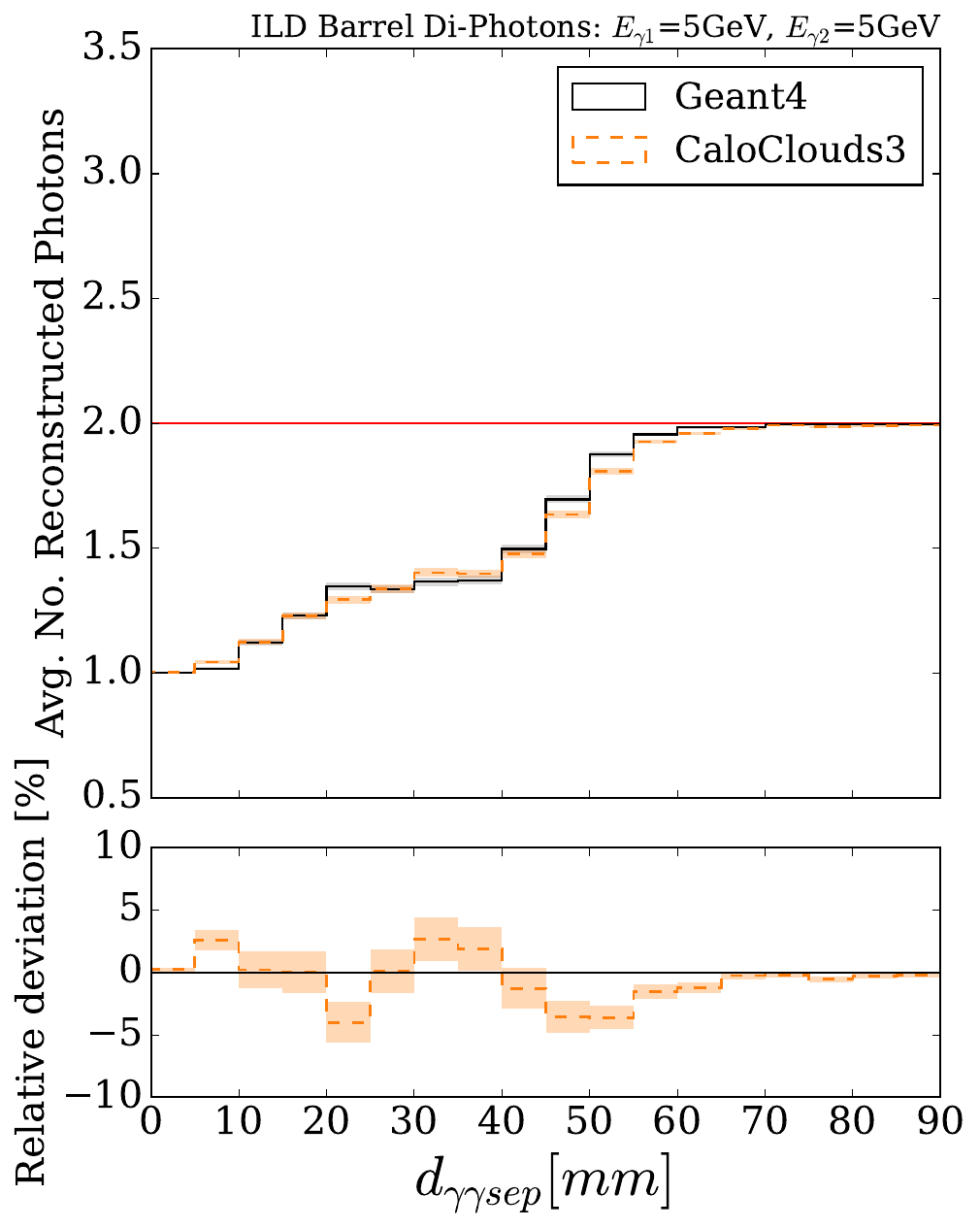}
    \includegraphics[width=0.3\textwidth]{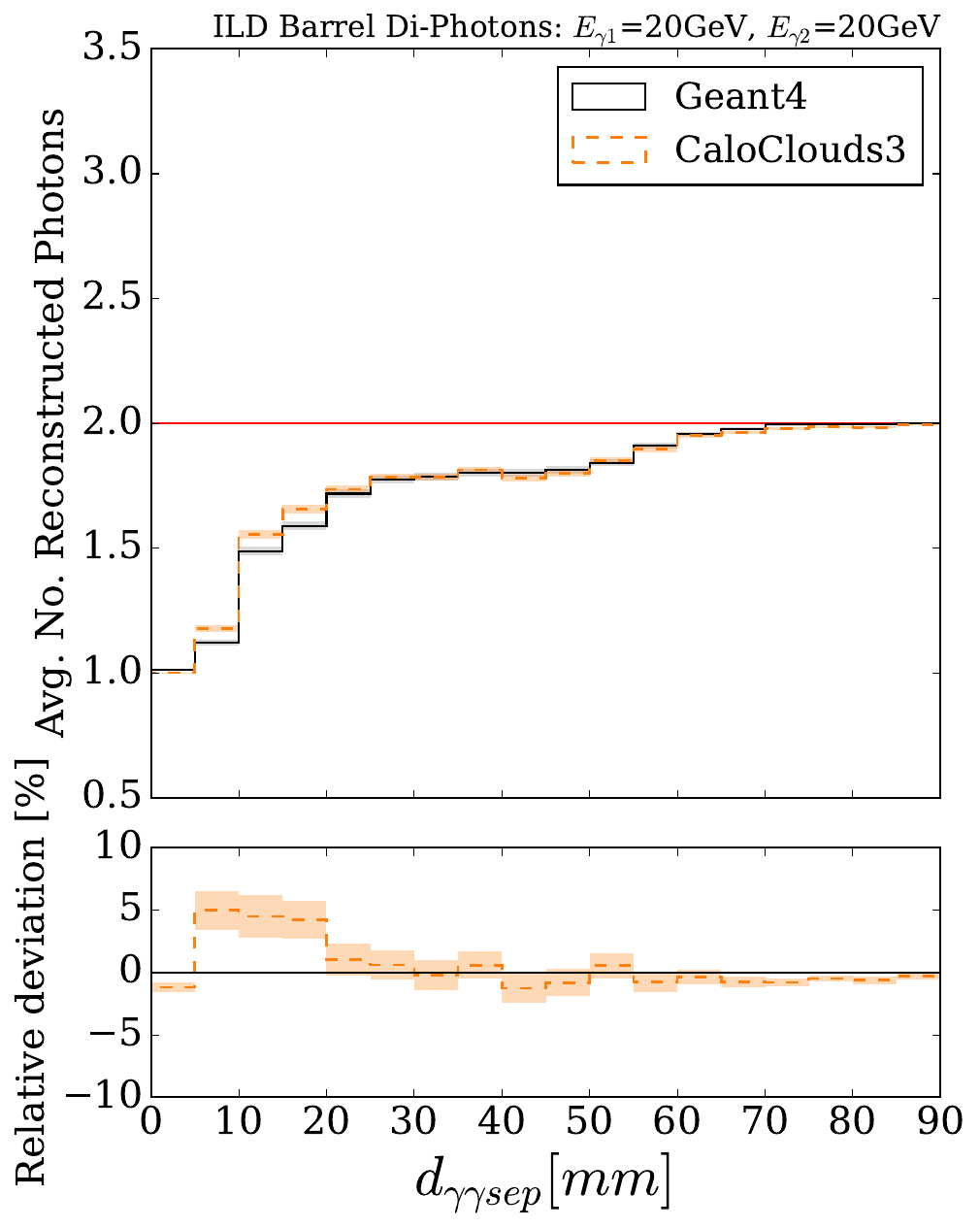}
    \includegraphics[width=0.3\textwidth]{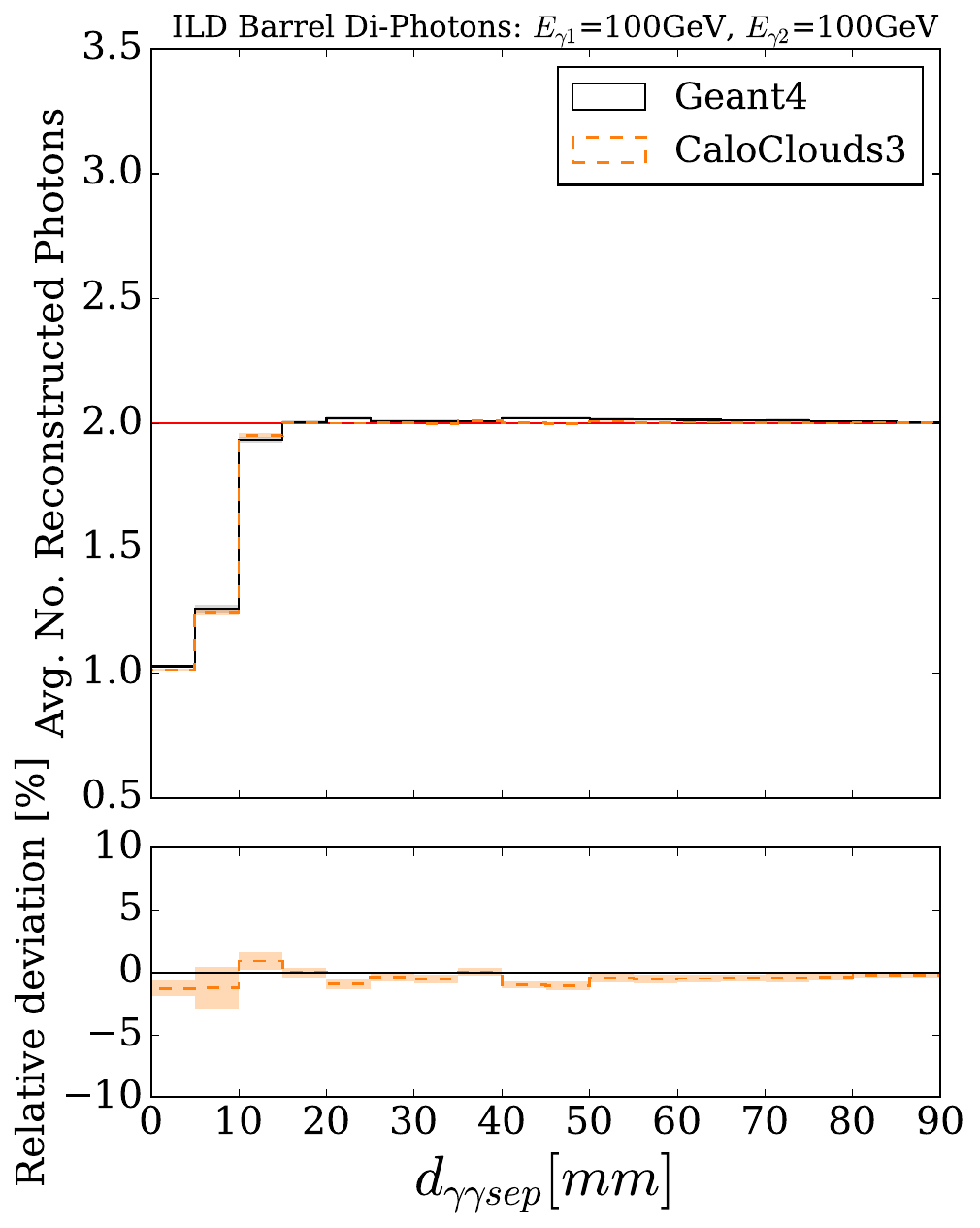}
    \caption{
        The reconstructed multiplicity of two photons fired into the same barrel segment, as modelled
        in \ccthree{} and \gfour{}.
        Three different energies are shown, from left to right: \(5~\si{GeV}\), \(20~\si{GeV}\) and \(100~\si{GeV}\).
        The shaded grey and orange bands in the top plot represent the errors of \gfour{} and \ccthree{} respectively, while the shaded orange band in the ratio plot below represents the propagated combined error on the ratio.
    }
\label{fig:diphot}
\end{figure}

In figure~\ref{fig:diphot}, we can see that \ccthree{} reproduces the behaviour of \gfour{} to within the fluctuations
expected from the errors.
In an analysis performed on samples generated by \ccthree{} we have confidence that the photon multiplicity would be
correctly reconstructed.

\section{Inference Speed}\label{sec:speed}

Faster inference than \gfour{} is a key objective. To fairly compare, we must run in the same software environment on the same hardware.

As detailed in section~\ref{sec:CC3arch}, as \ccthree{} has angular conditioning, it can be run in a realistic simulation toolchain.
Purely for timing purposes \cctwo{} has also been compiled into the \ddml{} stack; the angular conditioning is simply ignored.
We use the embedding of the two models in order to benchmark the inference speed, using showers only at orthogonal impact.

\begin{figure}[ht]
    \centering
    \includegraphics[width=0.5\textwidth]{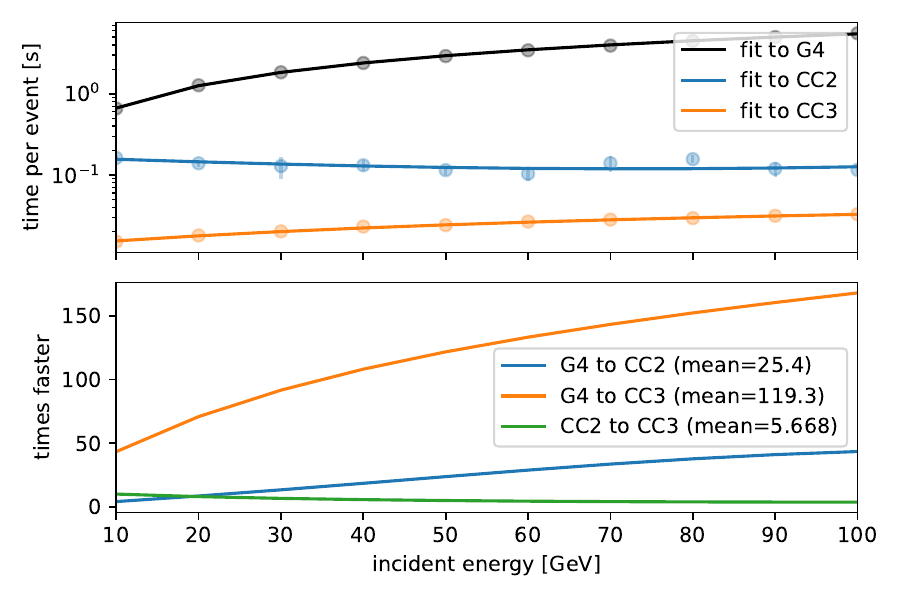}
    \caption{
        A comparison of the timing of \ccthree{} and \cctwo{} with respect to \gfour{}.
        The upper plot includes error bars for each point, calculated as the standard deviation of 3 runs,
        however only the error bars for \cctwo{} are large enough to be visible, the reasons for this are unknown.
        The lines in the upper plot are a linear fit, 
        and in the lower plot, the ratios of the fits are given.
        All timings are performed on a single thread of a single CPU, in the same cluster environment.
        In the legend, the mean speed increase across all incident energies is provided.
    }
\label{fig:timing}
\end{figure}

The results are shown in Figure~\ref{fig:timing}.
This timing is done on a compiled model and includes hit placement into the detector geometry.
Unfortunately, it is not currently possible to utilise a GPU in this setup, but this matches the current realistic use pattern, in which fast simulation is mainly done on CPU.
The CPU used for our tests was a AMD EPYC 7513 32-Core Processor
from DESY's Maxwell~\cite{desymaxwell} cluster.

From this comparison, we see that \ccthree{} is on average over \(6\) times faster than \cctwo{}, and two orders of magnitude faster than \gfour{}.
\gfour{} and \ccthree{} have roughly linear time complexity against the energy of the incident particles, which reflects the linear increase in number of points in the shower.
A significant portion of the time for a \cctwo{} shower is spent in the \showerf{} model, resulting in a large base time, even for low energy showers.
By reducing the \showerf{} model to the essentials, \ccthree{} has a much more economical base time and so continues to be advantageous at low energies. 
It is also seen that the variation on the inference time of \cctwo{} is significantly larger than that of \ccthree{} or \gfour{}. We did not anticipate this behaviour from \cctwo{}, and do not know what causes it.

\section{Conclusions}\label{sec:conclusions}

In this paper we have described the final iteration of the \caloc{} series, \ccthree{}.
Within the constraints of this architecture, the hyperparameters of this model provide fastest inference without compromising accuracy.
The improvements therefore facilitate more analysis work than otherwise possible,
or potentially a reduction in the carbon footprint of the simulation. 

By training the model in a location agnostic manner,
and conditioning on incident angle in addition to energy, this model is applicable across the barrel region of the whole detector.
It remains able to accurately recreate the kinematics of the photon shower with these changes,
and the reconstructed metrics from the \ccthree{} fast simulation are compatible with those of a \gfour{} simulation,
demonstrating that this model is ready to replace \gfour{} in a simulation chain.

Finally, this work demonstrates two broader concepts.
Firstly, representing data as a point cloud offers significant advantages in high granularity calorimeter simulations;
no other approach remains efficient in the sparse data space, while capturing the details of the high granularity calorimeter.
Secondly, as a great majority of the models used in physics were designed for natural image processing,
optimal hyper parameter choices may differ significantly from those typically used.

If the detector design itself were modified, we would expect ideal hyperparameters choices to also change, but in other respects the model is agnostic to the specifics.
In the same vein, this architecture could be applied to the end-cap, requiring only an additional dataset, and some minimal hyperparameter adjustments. 

In future work, we aim to apply these techniques to other common particles observed in the detector,
and to investigate the potential for reconstruction using the physics knowledge of a fast simulation model.

\begin{appendices}

\section{\cctwo{} Architecture}\label{apen:cctwo}

In this appendix, we briefly describe the penultimate version, \cctwo{}\footnote{Previously written as \textsc{CaloClouds II}}.
A complete description, along with a description of the training data, is available in the existing paper~\cite{Caloclouds2023}.

\begin{figure}[ht]
    \centering
    \includegraphics[width=0.5\textwidth]{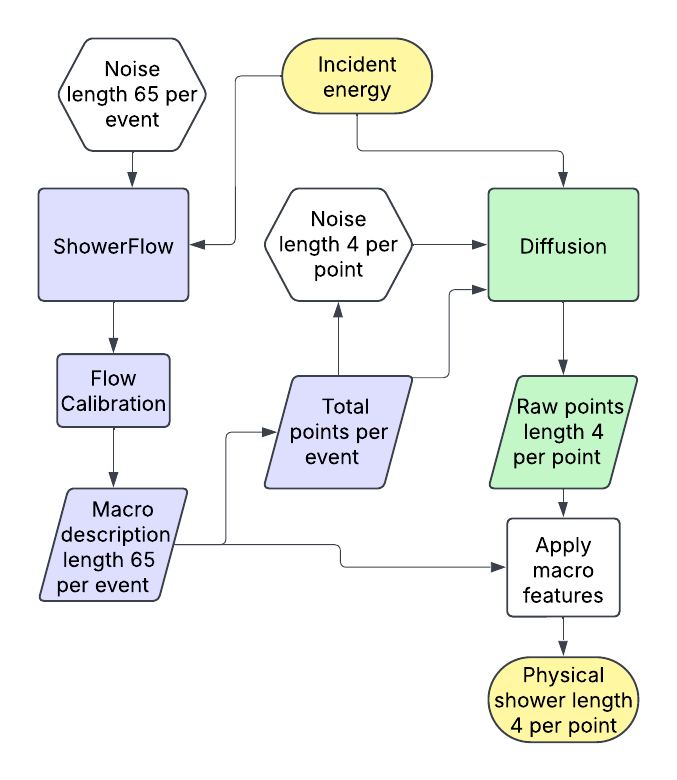}
    \caption{
        The model architecture of \cctwo{}.
        Parts belonging to \showerf{} are in purple, and those belonging to the diffusion model are in green.
        A brief description of the model is provided in~appendix~\ref{apen:cctwo},
        for further details see~\cite{Caloclouds2023}.
    }
\label{fig:CC2arch}
\end{figure}

The model itself is composed of a normalising flow, referred to as \showerf{}, (shown in purple in figure~\ref{fig:CC2arch}) and a distilled diffusion model (shown in green in figure~\ref{fig:CC2arch}).

\showerf{} finds the macro features of the shower, and the distribution along the shower axis.
It is conditioned on the incident energy of the photon, to produce a flow that transforms a
vector of Gaussian noise into a vector whose elements predict the following 65 variables:

\begin{itemize}
    \item[\(1\)\hspace{15pt}] The total number of points in the shower, rescaled so that the dataset is roughly between \(0\) and \(1\).
    \item[\(2\)\hspace{15pt}] The total visible energy of the shower, rescaled so that the dataset is roughly between \(0\) and \(1\), and clipped to a minimum of \(40~\si{MeV}\).
    \item[\(3\) to \(5\)\hspace{6pt}] The CoG of the shower in the \(x,y,z\) directions, rescaled to have a mean of \(0\) and standard deviation of \(1\).
    \item[\(6\) to \(35\)\hspace{1pt}] The number of points in each layer divided by the maximum number of points in any layer, such that all numbers are between 0 and 1 inclusive.
    \item[\(36\) to \(65\)] The energy in each layer, divided by the maximum energy in any layer, such that all numbers are between 0 and 1 inclusive.
\end{itemize}

This output is is used to calculate physical values to \emph{Apply macro features}. 
The diffusion model is then conditioned on the incident particle energy and the total number of points predicted by \showerf{}.
The conditioned diffusion model creates a set of raw points, drawn from an iid distribution.
This diffusion model has been distilled such that it generates points in a single network evaluation.
Distillation was not seen to cause any significant loss of accuracy as shown in~\cite{CaloClouds2}.

The raw points created by the diffusion model are rescaled to \emph{Apply macro features}.
Starting from the first detector layer the photon encounters and working outwards:
\begin{enumerate}
    \item The number of points predicted by \showerf{} for this layer is selected and
        moved to the correct position along the shower axis.
    \item The total energy of this layer is rescaled to match the energy predicted by \showerf{}.
\end{enumerate}

Once all layers are finished, the points are offset to match the CoG in \(x\) and \(y\) predicted by \showerf{}.
After projecting the energy into the detector cells, the occupancy of a shower generated by the model is not well matched to the \gfour{} result.
A correction factor is applied to counter this; this correction factor
modifies the number of hits requested from the diffusion model to using a polynomial function of the number of
hits predicted by \showerf{}.
Finally, one scalar correction factor for the energy of the points is applied.

\section{Further plots of all angles at fixed energies}\label{appendix3}

\begin{figure}[ht]
    \centering
    \includegraphics[width=0.9\textwidth]{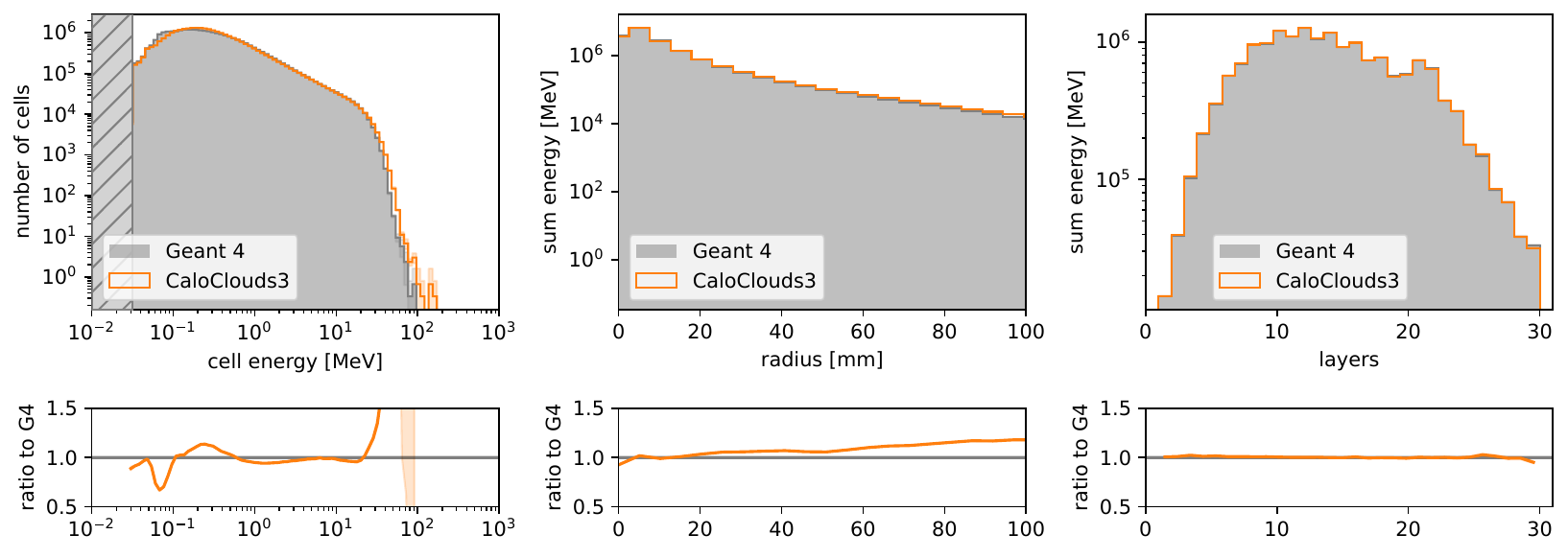}
    \caption{
        Energy spectrum of \ccthree{} at random angles, with all 10 energy increments combined.
        Layout is the same as in figure~\ref{fig:3key_0}.
    }\label{fig:3key_allang_0}
\end{figure}

\begin{figure}[ht]
    \centering
    \includegraphics[width=0.9\textwidth]{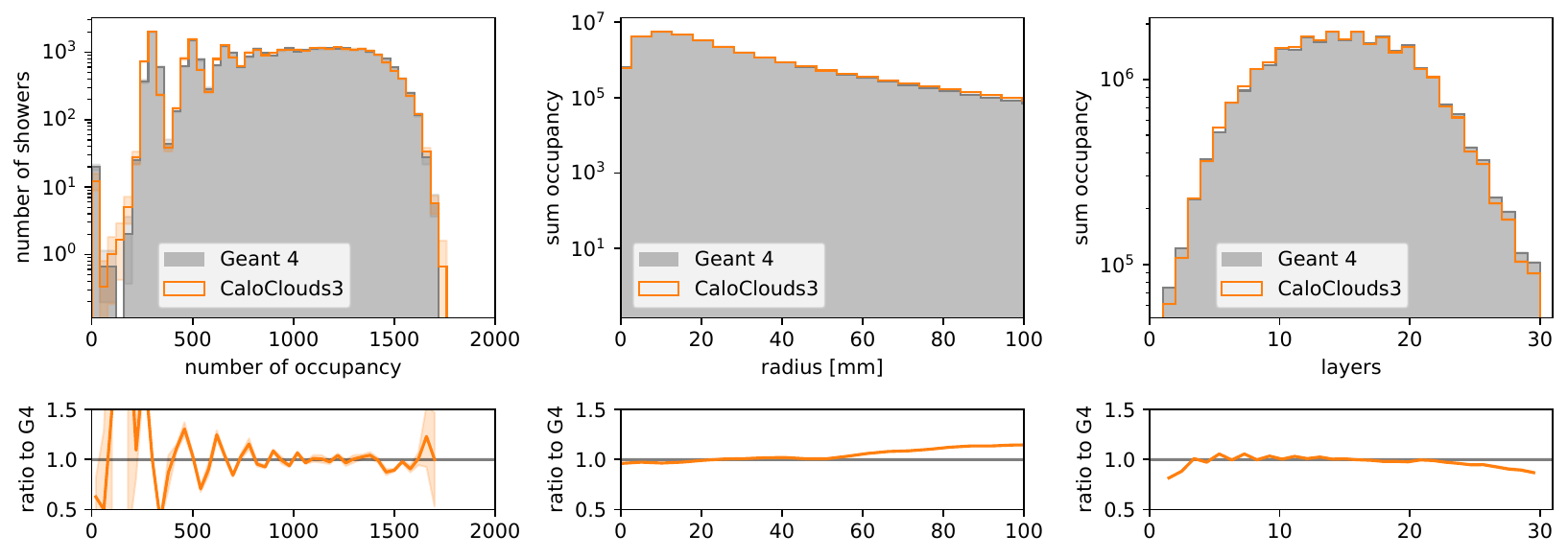}
    \caption{
        Occupancy spectrum of \ccthree{} at random angles, with all 10 energy increments combined.
        Layout is the same as in figure~\ref{fig:3key_1}.
    }
\label{fig:3key_allang_1}
\end{figure}

In section~\ref{sec:fixed_energies}, we show kinematic performance metrics for \ccthree{} at three fixed energies,
in a dataset with random angles.
For completeness, plots containing the combination of all 10 energies are presented in figure~\ref{fig:3key_allang_0} and figure~\ref{fig:3key_allang_1}.
It is from these histograms that the metrics in table~\ref{tab:fixed_energies_metrics} are calculated.

\section{Angular error in PCA}\label{appendix2}

In this section, we explore the structure of the error between the calculated internal angle and the true direction of the incident particle
as described in section~\ref{sec:angular_reconstruction}.
For each shower, an internal direction is calculated using PCA.
Either the calculated uses all the hits, or for the improved metric, only the \(4\%\) highest energy hits are used.
This can then be compared to the true direction of the incident particle, and the difference between the calculate and true direction is refered to as a bias.
For example, if the PCA direction has lower \(\theta\) than that of the incident particle, the bias in \(\theta\) will be negative.

\begin{figure}[ht]
\centering
\includegraphics[width=5.5in]{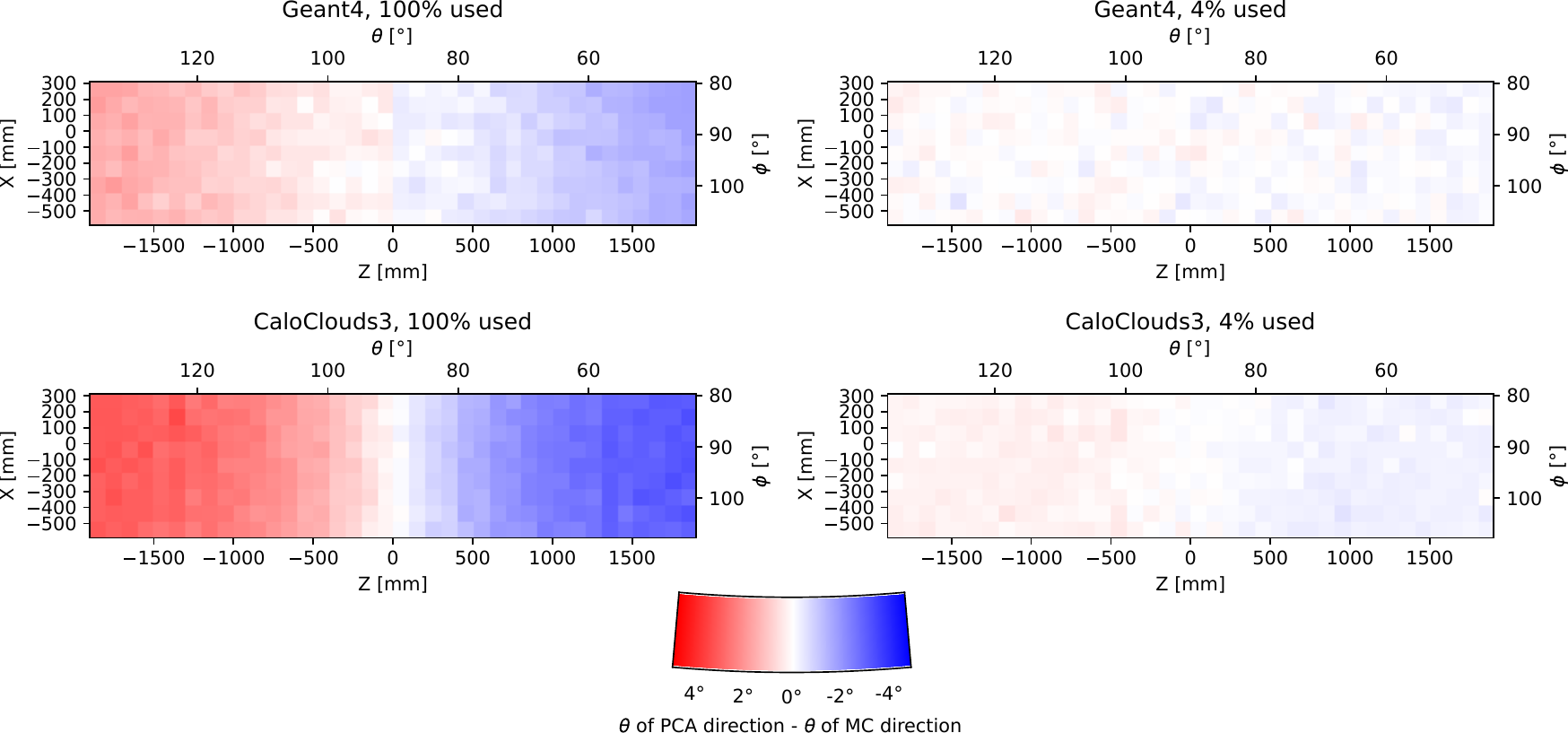}
\caption{Heatmaps of the bias from the \(\theta\) direction of the incident particle to the first PCA component.
         In the top row, the model generating the shower from which the PCA is calculated is \gfour{}, and in the bottom row the model is \ccthree{}.
         In the left column, \(100\%\) of the hits are used, and in the right column, \(4\%\) of the hits are used.
         The two distinct colour areas for \ccthree{}, with \(100\%\), create the two distinct peaks in figure~\ref{fig:angular_error_hist}.
        }\label{fig:theta_error_heatmap}
\end{figure}

\begin{figure}[ht]
\centering
\includegraphics[width=5.5in]{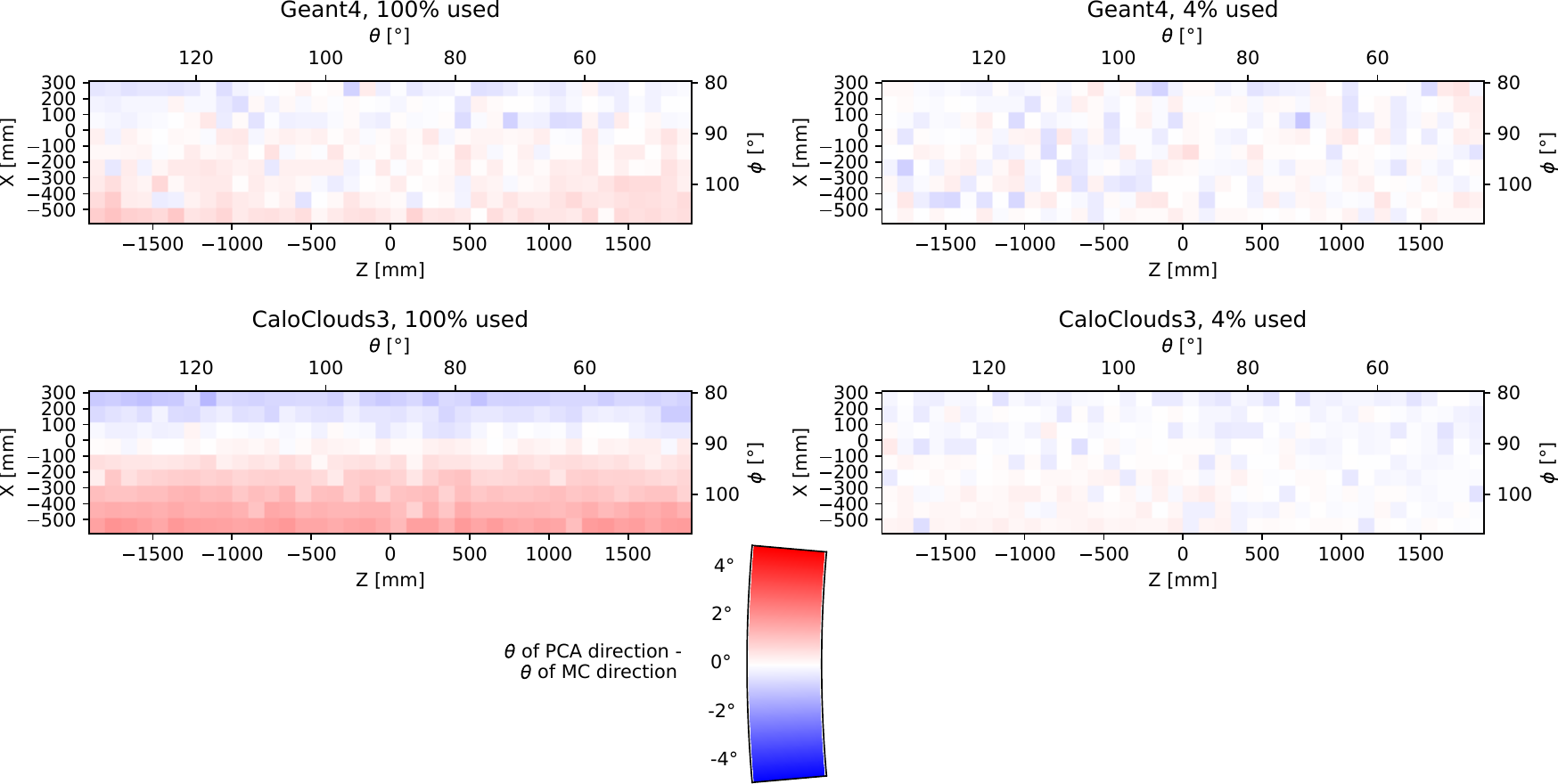}
\caption{Heatmaps of the bias the \(\phi\) direction of the incident particle to the first PCA component.
         In the top row, the model generating the shower from which the PCA is calculated is \gfour{}, and in the bottom row the model is \ccthree{}.
         In the left column, \(100\%\) of the hits are used, and in the right column, \(4\%\) of the hits are used.
         Given on the same colour scale as figure~\ref{fig:theta_error_heatmap}
        }\label{fig:phi_error_heatmap}
\end{figure}

Figure~\ref{fig:theta_error_heatmap} shows the bias in the \(\theta\) direction (polar angle) of the calculated internal angle, for both \gfour{} and \ccthree{}.
In the first column, where \(100\%\) of the hits are used, the bias is clearly symmetric;
in the positive \(\theta\) direction, the error is positive, and in the negative \(\theta\) direction, the error is negative.
This is a physical result of the secondary particles encountering more active material at shallower angles, and clipping of the shower at the calorimeter surface, both of which rotate the first PCA component away from the true direction of the incident particle.
This artifact is present in both \gfour{} and \ccthree{}, but significantly more pronounced in \ccthree{}, hence it has a double peak in figure~\ref{fig:angular_error_hist}.

From the second column, we see that when considering the PCA fit to only the \(4\%\) of highest energy hits, the effect is almost entirely eliminated.

In a similar way, bias in \(\phi\) can be plotted, shown in figure~\ref{fig:phi_error_heatmap}.
Here, in the left hand column we see similar angular effects, which bias the shower direction calculated by PCA towards shallower angles than the correct incident particle angle.
In the left hand column, with only the highest \(4\%\) of hits, we see the backscattering effect vanish.

\end{appendices}

\section{Acknowledgements}
We thank Dirk Zerwas for the valuable comments on the manuscript. This research was supported in
part by the Maxwell computational resources operated at Deutsches Elektronen-Synchrotron DESY,
Hamburg, Germany. This project has received funding from the European Union’s Horizon 2020
Research and Innovation programme under Grant Agreement No 101004761. We acknowledge
support by the Deutsche Forschungsgemeinschaft under Germany’s Excellence Strategy – EXC
2121 Quantum Universe – 390833306 and via the KISS consortium (05D23GU4, 13D22CH5)
funded by the German Federal Ministry of Education and Research BMBF in the ErUM-Data
action plan. A.K. has received support from the Helmholtz Initiative and Networking Fund’s initiative
for refugees as a refugee of the war in Ukraine. P.M. has benefited from support by the CERN Strategic R$\&$D Programme on Technologies for Future Experiments \cite{EPRD}.
Finally, we thank the anonymous referees from JINST for their through and thoughtful reading, which was a great assistance in polishing this work.
\clearpage{}

\bibliographystyle{JHEP}
\bibliography{Cited}

\end{document}